\documentclass[twocolumn]{aastex63}
\usepackage{amsmath}

\def\dem71{DEM L 71\,}

\usepackage[normalem]{ulem}
\usepackage[rightcaption]{sidecap}
\usepackage{graphicx} 
\graphicspath{ {images/} }
\usepackage{hyperref}
\submitjournal{ApJ}
\shorttitle{Ion Charge States in CMEs}
\shortauthors{Laming et al.}

\begin{document}

\title{The Evolution of Ion Charge States in Coronal Mass Ejections
}

\correspondingauthor{J. Martin Laming}
\email{j.laming@nrl.navy.mil}

\author[0000-0002-3362-7040]{J. Martin Laming}
\affil{Space Science Division, Code 7684, Naval Research Laboratory, Washington DC 20375, USA}

\author[0000-0001-8875-7478]{Elena Provornikova}
\affil{The Johns Hopkins University Applied Physics Laboratory, Laurel, MD 20723, USA}

\author[0000-0002-8747-4772]{Yuan-Kuen Ko}
\affil{Space Science Division, Code 7684, Naval Research Laboratory, Washington DC 20375, USA}

\begin{abstract}
We model the observed charge states of the elements C, O, Mg, Si, and Fe in the coronal mass ejections (CMEs) ejecta. 
We concentrate on ``halo'' CMEs observed  {\it in situ} by ACE/SWICS to measure ion charge states, and also remotely by STEREO 
when in near quadrature with Earth, so that the CME expansion can be accurately specified. Within this observed expansion, we integrate equations
for the CME ejecta ionization balance, including electron heating parameterized as a fraction of the kinetic and gravitational energy gain of the
CME. We also include the effects of non-Maxwellian electron distributions, characterized as a $\kappa$ function. Focusing first on the 2010 April 3
CME, we find a somewhat better match to observed charge states with $\kappa $ close to the theoretical minimum value of $\kappa = 3/2$, implying a hard spectrum of non-thermal electrons. 
Similar, but more significant results come from the 2011 February 15 event, although it is quite different in terms of its evolution. We discuss the implications
of these values, and of the heating required, in terms of the magnetic reconnection Lundquist number and anomalous resistivity associated with CME evolution close to the Sun.

\vspace{0.5true in}

\end{abstract}

\section{Introduction}
Coronal Mass Ejections (CMEs) are arguably the most important
component of space weather. These extreme events of solar activity can
efficiently accelerate energetic particles and produce geomagnetic storms.
Understanding the eruption process of CMEs, in particular  physical
mechanisms that heat plasma and accelerate particles in the early stages of
CME evolution, is crucial to understanding and forecasting the CME component of space weather.

Ion charge states detected {\it in situ} at 1 AU provide crucial insight into the heating and evolution
of coronal mass ejections (CMEs) near the Sun. Different ionization stages of different elements ``freeze-in''
at different heights in the solar atmosphere, typically within about 6 $R_{\sun}$,  \citep[though denser material in prominences can freeze-in beyond this distance][]{rivera19}, and no longer change as electron density rapidly
decreases with height. This allows one to reconstruct the various episodes of heating and expansion.
Some of these charge states are uniquely produced by magnetic reconnection associated with CMEs, where high charge
states indicate the presence of high electron temperatures, and so offer a novel observational
approach to understanding this process, particularly its role in electron heating and acceleration. 

Our analyses of CME ejecta ion charge states began with an initial exploration in \citet{RLL07},
a spheromak model for a CME by \citet{RLL11}, and
breakout and flux-cancellation modeling of CMEs by \citet{lynch11}, while \citet{lepri12}
revealed spatial distribution of heating in the 2005 January 20 CME observed by both the Advanced Composition Explorer
(ACE) and Ulysses. These works quantified thermal energy input during early CME evolution and showed the importance of magnetic reconnection for CME heating, but many questions still remain to be answered. In particular, many details of the reconnection process resulting in different charge states distributions for CMEs of different speeds remain to be understood. Besides energy partitioning between bulk kinetic energy, magnetohydrodynamic (MHD) waves, and heat, details of the electron heating and acceleration
remain to be validated observationally. 

In this work we revisit the modeling of CME ejecta charge states with a considerably improved model, incorporating the effects of
non-Maxwellian distribution functions. Our CMEs are carefully chosen so that the expansion and morphological evolution can be properly specified. 
Integration of equations for time-dependent ionization balance within this evolution for different degrees of electron heating, and for different
electron distribution functions (taken here as $\kappa$ distributions), reveals a preference for non-Maxwellian electron distribution with plausible 
degrees of heating. We further discuss these results in terms of modeling of the electron distribution from the Fokker-Planck equation, and the
parameters such as Lundquist number, anomalous resistivity and plasmoid instability required. Section 2 below summarizes the theoretical context 
for this work. Section 3 outlines the time dependent ionization balance calculations and implementation of the non-Maxwellian electron distributions.
Section 4 introduces the data and CME events we study, while section 5 describes the results of our modeling. Section 6 continues with a discussion of implications of the results. Section 7 presents supporting MHD simulations of magnetic reconnection with multiple plasmoids. Section 8 presents the conclusions of the study.

\section{Theoretical Context}
\subsection{Plasmoid Instability}
While the model CME eruption geometries differ, most scenarios invoke
magnetic reconnection above and/or below the ejected plasma. Reconnection
below the CME may impart kinetic energy to the CME, and 
is arguably the most likely source of heat for the CME plasma and hence the high Fe charge states detected {\it
in-situ}. \citet{murphy11} discuss a range of other possible heating mechanisms
and are able to reject all of them except for heating by small-scale magnetic reconnection in the CME
current sheet.

However, it is widely believed that the fundamental component of a CME is a
magnetic flux-rope \citep{chen03,linton09b,liu10, Vourlidas2013}, which may arise independently of the reconnection, or
be formed by the plasmoid instability in high Lundquist number reconnection.
Simulations and theory are now starting to clarify exactly how
magnetic reconnection can heat or accelerate electrons and ions in this manner. In recent studies of magnetic
reconnection, the role of the secondary plasmoid instability of the
highly elongated reconnection current sheets
has been under investigation \citep{Shibata01,Bhattacharjee10,Cassak09, Landi2015, arnold21}.

Plasmoid (or magnetic island) formation is important to this work as a means of electron heating
\citep{drake06} and as a constraint on the rate of reconnection.
\citet{Loureiro07} have shown that
the criteria for onset of the plasmoid instability is that the plasma
Lundquist number $S\equiv 4\pi L V_A/(\eta c^{2})\gtrsim 10^4$, where $L$ is the length of the
current sheet, $V_A$ is the Alfv\'en speed in reconnection inflow and $\eta$ is the plasma resistivity.  This implies that
slow Sweet-Parker like reconnection in the regime of large $S$ never
happens because the current sheet is always disrupted into plasmoids. \citet{pucci14} 
argue that the ``ideal'' tearing mode takes over when the current sheet reaches a thickness
$\delta/L \sim S^{-1/3}$, which is by two orders of magnitude thicker than the
Sweet-Parker scaling $\sim S^{-1/2}$. \citet{uzdensky16} give a slightly different criterion, and
\citet{comisso16} provide a more general theory of plasmoid formation. A number of simulations by
\citet{Bhattacharjee10}, \citet{Lapenta08}, \citet{Daughton09}, and \citet{Loureiro2012}
demonstrated plasmoid generation and $S$-independent or weakly
dependent reconnection rates for $S > S_{crit}\sim 10^4$,  where $S_{crit}$ is the critical Lundquist number.
\citet{uzdensky10} show that the reconnection rate in the plasmoid
regime ($S>10^{4}$) is fast and independent of $S$, $cE_{eff}\sim S_{crit}^{-1/2}V_{A}B_{0}$.
However plasmoid generation in three dimensions may be more complex
\citep{Daughton11,WyperPontin2014}. Plasmoid formation, ejection and merging has been observed in CMEs by \citet{song12}, and in 
laboratory experiments by \citet{Hare17, Suttle2016}. The reconfiguring of the solar corona by reconnection during a CME is studied
by \citet{vandriel14}.

\subsection{Energy Partitioning}
Currently no strong prediction exists for how much plasma
heating should be expected as a function of CME kinetic energy. In the case of laminar (i.e. Sweet-Parker)
reconnection, \citet{priest14} gives an estimate that thermal energy released should be equal to the kinetic energy
in the incompressible case, increasing as $2\rho _o/\rho _i -1$ in compressible reconnection where $\rho _{i,o}$ are the 
densities of the inflow and outflow respectively. 

In the case of reconnection diffusion region restricted to a length $L_i$ of a magnetic field reversal extending
over length $L_e$ (see Fig. 1), \citet{priest14} gives
\begin{equation}
\frac{L_i}{L_e} = \frac{\rho _i^{1/2}}{\rho _e^{1/2}}\frac{1}{S_e}\frac{1}{M_e^{1/2}M_i^{3/2}} =
\frac{\rho _e^{1/4}}{\rho _i^{1/4}}\frac{1}{S_e}\frac{1}{M_e^2}
\end{equation}
where a factor involving the densities has been reinstated, $M_{e,i}=v_{e,i}/V_{Ae,i}$ are Mach numbers in terms of the
Alfv\'en speeds at exterior and interior inflow regions, and $S_e=4\pi L_eV_{Ae}/\eta c^2$ is the exterior Lundquist number. The second equality comes from
taking $B_i\sim B_e$, $v_i\sim v_e$ so that $M_{e}/M_{i}\sim \sqrt{\rho _e/\rho _i}$, the outflow kinetic
energy as a fraction of the inflow magnetic energy is 
\begin{equation}
\frac{\rm outflow~KE}{\rm inflow~magnetic~energy}=\frac{1}{2}\left(\frac{\rho _e^{1/4}\rho _i^{3/4}}{\rho _o}\right)\frac{1}{S_eM_e^2}.
\end{equation}
The remaining energy is taken to be heat, giving a ratio
\begin{equation}
{\frac{\rm thermal~energy}{\rm outflow~KE}}=2S_eM_e^2\frac{\rho _o}{\rho _e^{1/4}\rho_i^{3/4}}-1.
\end{equation}
Adopting $S_e\sim 10^4$ as the threshold for plasmoid instability implies $M_e\sim 10^{-2}$ and 
\begin{equation}
{\frac{\rm thermal~energy}{\rm outflow~KE}}=2r\left(\frac{\rho_i}{\rho_e}\right)^{1/4}\frac{S_e}{10^4}-1
\label{enfrac}
\end{equation}
where $r=\rho _o/\rho _i$. If $M_i>\beta$, a slow mode shock can develop in the inflow to the diffusion region, leading to $1<r<2.5$.
The kinetic energy comes only from the $L_i$ diffusion regions, while the magnetic islands that separate them contribute only heat. These
arguments do not account for any energy in waves \citep[c.f.][]{kigure10}. We anticipate that Alfv\'en waves should be part of the kinetic energy
budget, while magnetosonic waves are part of the thermal energy. Stronger compression leads to more heat, as in \citet{Provornikova16,Provornikova18}.

\begin{figure}[t]
\vspace{-1.5truein}
\centerline{\includegraphics[width=3.5in]{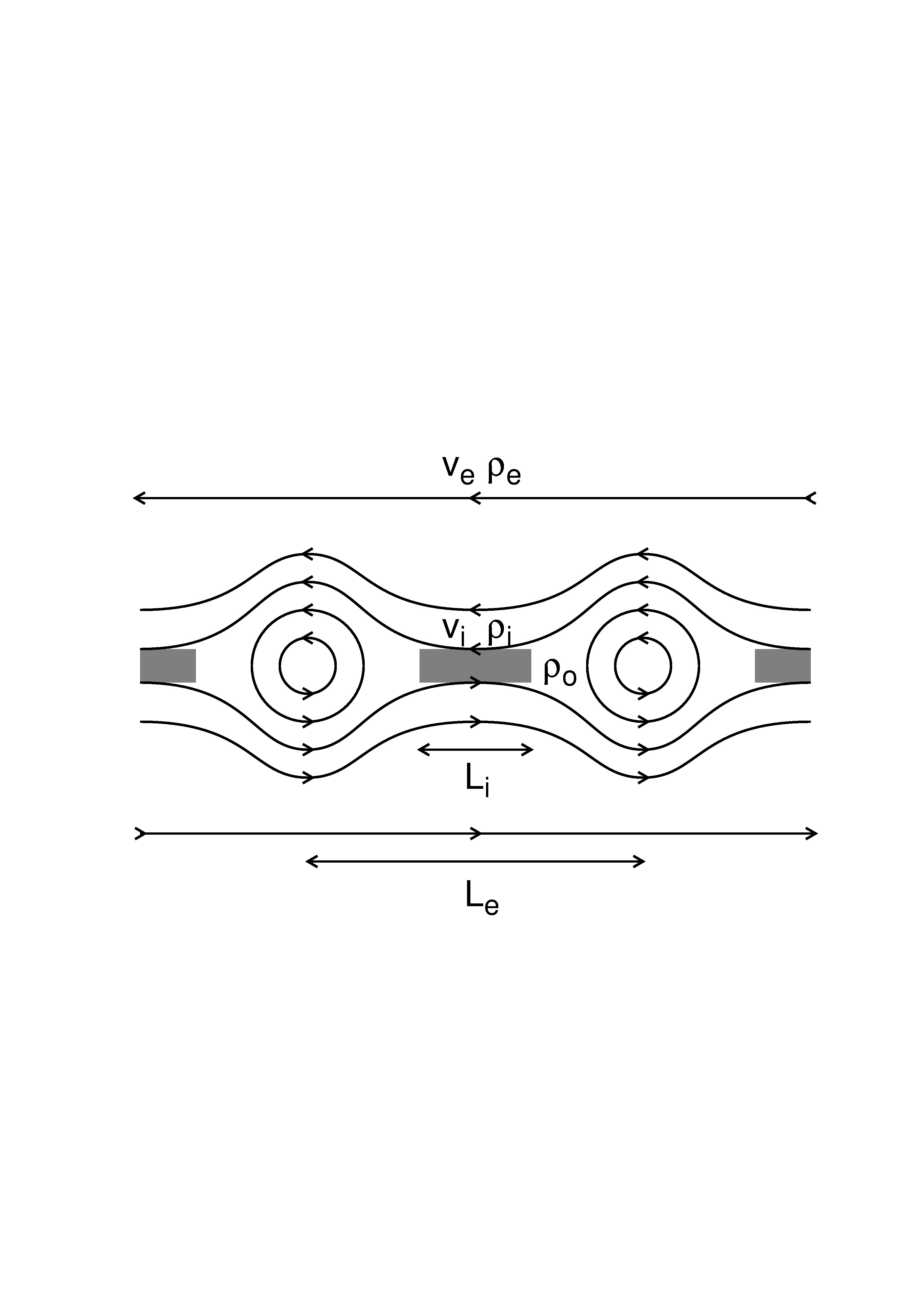}}
\vspace{-1.5truein}
\caption{Schematic diagram of a reconnection current sheet with characteristic external scale length $L_e$, and interior length of
diffusion region (shown in gray) of $L_i$. The exterior and interior flow velocities are $v_e$ and $v_i$ respectively. Magnetic islands
form either side of the diffusion regions, and are the sites of electron heating as they evolve and merge. \label{plasmoid}}
\end{figure}

An important feature of this heating will be the partitioning of energy between electrons and ions.
A large fraction of the released magnetic energy in flares may be channeled into high energy electrons, protons
\citep{lin71,emslie04,emslie05} and other ion species \citep{mason07}, as well as waves and other forms of kinetic energy.
The X-ray emission spectra reveal that energetic electrons in flares typically consist of a hot component in the range
of 20-30 keV with higher energy powerlaw tails \citep{holman03}. Over-the-limb observations of flare energy release
reveal that almost all electrons in the flaring site can be accelerated to
became part of the energetic component, whose pressure approaches that
of the magnetic field \citep{krucker10}. \citet{glesener2013} suggest that flare-accelerated
electrons collisionally deposit their energy into plasma heating in the CME to at least 11 MK.

\subsection{Anomalous Resistivity} \label{secAR}
Observationally, conditions in CME current sheets may be closer to the
threshold for plasmoid generation than are usually thought.
Observational signatures of drifting and pulsating structures in solar
flare associated current sheets have been interpreted by
\citet{Kliem00} and \citet{Karlicky04} as evidence of embedded
secondary plasmoids. \citet{takasao12} also reported EUV observations of
plasmoid emergence, collisions and injections in the rising phase of the limb flare.
For typical coronal parameters
($L\sim 10^8$ cm, $V_A\sim 10^8-10^{10}$ cm s$^{-1}$, $\eta\sim 10^{-16}$ s$^{-1}$),
$S$ is about $10^{12}-10^{14}$, and is plausibly even larger at a large scale current sheet with larger
$L$ behind a CME. However these estimates assume a Spitzer electrical resistivity. A more
accurate idea of the Lundquist number may come from using an anomalous resistivity, increased
from the classical (i.e. Spitzer) value. Anomalous resistivity is
thought to arise in conditions where plasma turbulence can scatter
current carrying electrons, and can do so more effectively than the
Coulomb collisions between electrons and ions \citet{lazarian99}.
It may be estimated by replacing the isotropization frequency by
Coulomb collisions with the relevant wave-particle interaction rate.

\citet{Wu09} give quiescent values of the anomalous resistivity of
$\times 10^7$ the Spitzer value, with transient values as high as
$\times 10^9$ in a background plasma temperature of $10^7$ K using
solutions of the one dimensional Vlasov equation designed to model the
Buneman instability in a reconnecting current sheet, and therefore
model electrons in Langmuir turbulence. \citet{lin07} estimate a much higher resistivity from the observed width of current sheets
trailing CMEs, as high as $10^{12}$ times the classical resistivity.

Incorporating such an anomalous resistivity brings the Lundquist number
down to $S\sim 10^5$, much closer to the threshold for the production of magnetic islands
of about $10^4$. Thus the increasing Lundquist number with CME speed in this range much more
plausibly suggests increasing production of magnetic islands within the current sheet, such
that the associated electron acceleration can become more important as the CME speed increases.

\subsection{Electron Acceleration}
Models of electron acceleration in
reconnecting current sheets behind the erupting CME make definite
predictions about the non-Maxwellian electron distribution
\citep[e.g.][]{drake06,liu08, Guo16}. \citet{drake06} give analytic and
numerical treatments. In the weakly driven case, where the back reaction of the accelerated electrons can be neglected, the power law index
of the electron distribution at high energies can be related to the
geometry of the magnetic islands formed in the current sheet. In strongly
driven cases, this dependence is lost, and the electron plasma
$\beta=8\pi n_ekT_e/B^2$ plays a more important role in limiting the
acceleration, in the sense that higher plasma $\beta$ restricts the energy acquired by the
electrons. When the back pressure of the accelerated particles approaches that of the magnetic
field, island contraction is quenched and particle acceleration ceases.
\citet{liu08} give numerical simulations in the test
particle limit, and find similar electron spectra to
\citet{drake06}.
\citet{oka10JGRA,oka10ApJ,Karlicky11} also explore the electron
heating in plasmoids with different explanations for the
acceleration. Stronger electron acceleration is also found in magnetic islands with
strong compression, resulting from the reconnection of relatively unsheared field lines and
hence with a weak guide field \citep{guidoni16,li18,Provornikova16}.

More recently, \citet{drake13} and \citet{montag17} have given Fokker-Planck treatments of
electron acceleration by a Fermi mechanism in coalescing magnetic islands.
\citet{montag17} incorporate the effects of compressibility and
a guide field. Their incompressible limit recovers the result of \citet{drake13},  but compressible
plasma allows for harder electron spectra, and hence more energy going into the electrons. Lower ambient
plasma $\beta$ would lead to potentially more magnetic field destruction, and hence higher plasma
compression in the current sheet and stronger electron acceleration, reinforcing the idea
\citep{drake06, Provornikova16, Provornikova18} that lower ambient plasma $\beta$ leads to stronger heating. The
high energy power law component does not extend to infinite energy, but is limited by the time
available for particle acceleration. The presence of a guide field also limits the degree of
compression in the current sheet, and further limits the energy in electrons.

\citet{arnold21} study electron acceleration using an MHD model of reconnection incorporating fluid ions and electrons, as well as particle electrons to model the acceleration process. As the Lundquist number increases in the range $1.2\times 10^7$ to $6.1\times 10^9$ the electron
spectra become harder, but only with electron velocity power law index decreasing from about 3.3 to 3. However, \citet{arnold21} vary the Lundquist number through the system size, not the magnetic field or plasma $\beta$. Larger systems produce more reconnected flux and hence more extended power laws.

\section{Ionization and Recombination of CME Ejecta}

The heating and acceleration of electrons during reconnection through
magnetic islands in current sheets behind erupting CMEs in the corona will have direct
observational consequences in the ion charge states measured at 1 AU.
The charge states evolve through ionization and
recombination as a function of temperature and density up to heights
of 3 to 6 solar radii heliocentric distance but are ``frozen in''
thereafter. Thus {\it in situ} measurements of ion charge state distributions
hold a unique potential for diagnosing the conditions throughout the CME
eruption.

We model the charge states within the CME ejecta for a variety of
ions using an adaptation of the BLASPHEMER (BLASt Propagation in
Highly EMitting EnviRonment) code
\citep{laming02,laming03b,laming03c}, which given the temperature and
density history of a Lagrangian plasma parcel determines the ionization balance as
it expands in the solar wind. The density $n_{iq}$ of ions of element
$i$ with charge $q$ is given by
\begin{eqnarray} &\frac{dn_{iq}}{dt} &=
n_e\left(C_{ion,q-1}n_{i~q-1}-C_{ion,q}n_{iq}\right)+\\
\nonumber &&n_e\left(C_{rr,q+1} +C_{dr,q+1}\right)n_{i~q+1}- n_e\left(C_{rr,q}+
C_{dr,q}\right)n_{iq}
\label{ionden}
\end{eqnarray}
where $C_{ion,q}, C_{rr,q}, C_{dr,q}$ are the rates for electron
impact ionization (including autoionization following inner-shell excitation), radiative recombination and dielectronic
recombination respectively, out of the charge state $q$. These rates
are the same as those used in the ionization balance
calculations of \citet{bryans06}, with more recent updates given in
\citet{RLL07} and \citet{laming20}. The electron density $n_e$ is
determined from the condition that the plasma be electrically
neutral. The ion and electron temperatures, $T_{iq}$ and $T_e$ are
coupled by Coulomb collisions by
\begin{eqnarray} \nonumber \frac{dT_{iq}}{dt}&=& -0.13n_e\frac{\left(T_{iq}-T_e\right)}{M_{iq}T_e^{3/2}}
\frac{q^3n_{iq}/\left(q+1\right)}{\left(\sum _{iq}
n_{iq}\right)}\left(\frac{\ln\Lambda}{37}\right)\\
&& -\frac{4}{3}\frac{\gamma
_{iq}U_w}{n_{iq}k_{\rm B}}
\label{iontemp}
\end{eqnarray} and
\begin{eqnarray}
\nonumber \frac{dT_e}{dt}&=& \frac{0.13n_e}{T_e^{3/2}}\sum
_{iq}\frac{\left(T_{iq}-T_e\right)}{M_{iq}}
\frac{q^2n_{iq}/\left(q+1\right)}{\left(\sum _{iq}
n_{iq}\right)}\left(\frac{\ln\Lambda}{37}\right) \\
&& -\frac{T_e}{
n_e}\left(\frac{dn_e}{dt}\right)_{ion} - \frac{2}{3n_ek_{\rm B}}
\frac{dQ}{dt} +\sum _{iq}\frac{4}{3}\frac{\gamma _{iq}U_w}{n_ek_{\rm
B}}.
\label{etemp}
\end{eqnarray}
Here $M_{iq}$ is the atomic mass of the ions of element $i$ and charge
$q$ in the plasma, and $\ln\Lambda\simeq 28$ is the Coulomb
logarithm. The first term on the right hand side of each equation represent temperature
equilibration by Coulomb collisions. The term in $dQ/dT$ in the electron temperature
equation represents plasma energy losses due to
ionization and radiation. Radiation losses can be taken from
\citet{summers79}. They are generally negligible here. The term
$-\left(T_e/n_e\right)\left(dn_e/dt\right)_{ion}$ gives the reduction
in electron temperature when the electron density increases due to
ionization.
Recombinations, which reduce the electron density, do not
result in an increase in the electron temperature in low density
plasmas, since the energy of the recombined electron is radiated away
(in either radiative or dielectronic recombination), rather than being
shared with the other plasma electrons as would be the case for
three-body recombination in dense plasmas.
The terms in $\gamma
_{iq}U_{\omega}$ represent heating of individual plasma components by
turbulence, as opposed to bulk heating of the plasma itself as is
usually assumed. Electron or ion heating can be modeled with growth or damping rates
($\gamma _{iq}$) and turbulent energy densities ($U_{\omega}$)
calculated from quasi-linear plasma theory. Here we simply put $\gamma
_{iq}U_{\omega}$ for electrons only equal to a fraction of the gain in kinetic and gravitational energy of the expanding CME.

The background densities and temperatures evolve according to the observed hydrodynamic expansion, in an operator splitting method. The expansion is specified by observations described in Section 4 and described in more detail in Section 5. The density varies as
\begin{equation}
    n\propto \left\{\begin{array}{ll}
    1/\left(r-R_{\sun}\right)^2v_{exp}, & {\rm for} \quad r\le r_{\rm Alf}\\
    1/r^3, &{\rm for}\quad r > r_{\rm Alf}
    \end{array}\right\}
    \label{denseq}
\end{equation}
which captures the expansion at increasing velocity $v_{exp}$ from an explosion at a point on the solar surface for heliocentric radii, $r$, within the solar wind/CME Alfv\'enic surface, $r_{\rm Alf}$. For $r > r_{\rm Alf}$ an
approximate $1/r^3$ expansion takes over once the solar wind is essentially hydrodynamic, no longer accelerating, and
the plasma $\beta > 1$. In this regime $T\propto n^{2/3}$ as per adiabatic expansion. The starting density is determined from {\it in-situ} observations at 1 AU extrapolated back to the solar corona.

Extension of the non-equilibrium ionization modeling to incorporate a
non-Maxwellian electron distribution as predicted by for example
\citet{drake06} is a key part of this work. Other electron heating
mechanisms are not expected to energize the {\em whole} population of
electrons, as the Fermi acceleration in magnetic islands is expected
\citep{drake06} and observed \citep{krucker10} to do. We use expressions
for power law index analytically derived by \citet{drake06} and \citet{montag17}
to connect to magnetic island geometry. For $F\propto v^{-\gamma}$
\citet{montag17} find
\begin{equation}
\gamma = \frac{3}{2} -5\frac{\nu}{R^2}\frac{\dot{n}}{n}
+\sqrt{\left(\frac{3}{2} -5\frac{\nu}{R^2}\frac{\dot{n}}{n}\right)^2+30\frac{\nu}{R^2}\frac{c_{Aup}}{L}}.
\end{equation}
Here $\nu =\left( \Omega _e\pi/4\right) \delta B^2\left(k_{||}=\Omega _e/v_{||} \right)/B^2$ is the electron pitch angle scattering rate in
terms of the electron gyrofrequency, $\Omega _e$, and the magnetic field fluctuations $\delta B$ at parallel wavevector
$k_{||}=\Omega _e/v_{||}$.
In other terms $R=2\dot{n}/3n -\dot{B}/B$ ($n$ is the plasma density and $B$ is the background magnetic field)
and $c_{Aup}$ is the upstream Alfv\'en speed. In incompressible
conditions, $\dot{n}=\partial n/\partial t =0$ and equation 9 reduces to the result of \citet{drake13}. A similar expression is found by \citet[][their equation 65]{leroux15}.

The index depends on
the degree of plasma compression and pitch angle scattering frequency in
plasmoid dominated reconnection. The latter depends on the intensity of waves
produced by reconnection and their reflection near the current sheet.
Plasmoid formation, merging, oscillation and ejection during reconnection
process inevitably produce waves propagating away from the reconnection site.
Frequencies and wavelengths of generated waves are related to temporal and
spatial scales of plasmoid dynamics. \citet{Provornikova18} showed that
due to gradients in magnetic field and plasma density near the current sheet,
magnetosonic waves produced in reconnection undergo reflection back towards
the current sheet, increasing the pitch angle scattering, and hardening the
electron spectrum. Reflection is more efficient in low-$\beta$ plasma leading to increased particle acceleration. Wave frequencies
would be determined by the dimensions of magnetic islands in the outflow. In
such conditions, we should expect strong particle acceleration. 

\begin{figure}[t]
\centerline{\includegraphics[width=3.5in]{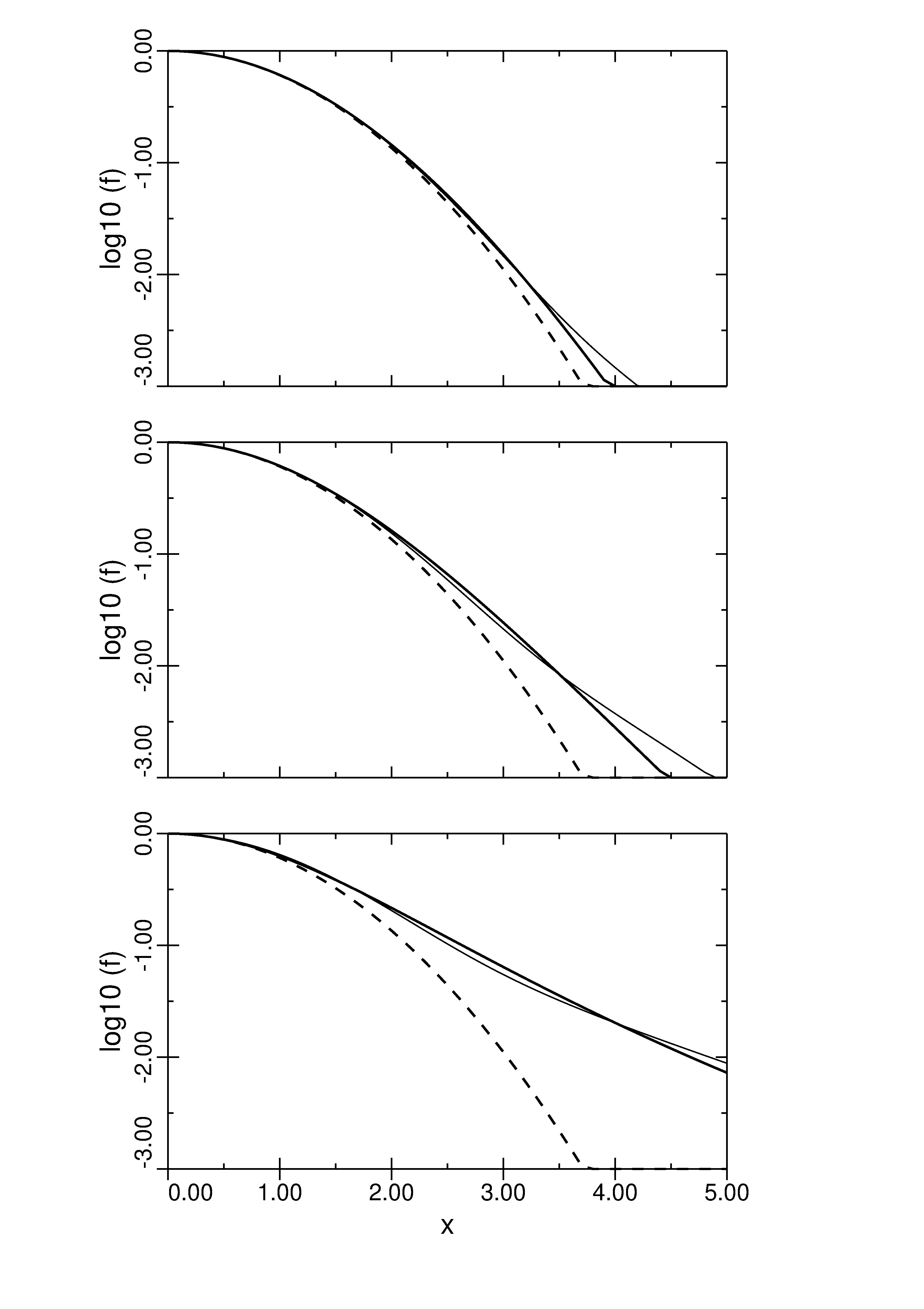}}
\caption{Examples of the full $\kappa$ distribution (thick solid), the underlying Maxwellian distribution (thick dashed) and the approximate
$\kappa$ function expressed as a sum of Maxwellians (thin solid) for $\kappa =$ 30 (top), 
10 (middle) and 3 (bottom) as a function of $x=v/v_t$. \label{hifibk}}
\end{figure}

\begin{table*}[t]
\begin{center}
\caption{2010 April 3 CME Off-Limb Plasma Parameters}
\begin{tabular}{cccccc}
\hline Radius ($R_{\sun}$) & $V_A$ (km s$^{-1}$)  & $\int V_Adl$ & $n_e$& $S=\int V_Adl\sigma/4\pi c^2$ & $S^{\prime}$ \\\hline
1.1 & 1000 & 100 &$3\times 10^9$  & $6\times 10^{11}$ & $1\times 10^4 - 3.6\times 10^7$ \\
1.5 & 300& 220   &$2\times 10^8$  & $1\times 10^{12}$&  $8\times 10^3 - 2.4\times 10^7$\\
2.0 & 270& 355& $3\times 10^7$ & $2\times 10^{12}$& $ 6\times 10^3 - 1.8\times 10^7$\\
2.5 & 250& 480& $1\times 10^7$ & $2.5\times 10^{12}$& $5\times 10^3 - 1.5\times 10^7$\\
3.0 & 200& 580& $6\times 10^6$ & $3\times 10^{12}$& $4\times 10^3 - 1.2\times 10^7$\\
5.0 & 200& 980& $1\times 10^6$ & $5\times 10^{12}$&  $3\times 10^3 - 1\times 10^7$\\
\hline \end{tabular}
\end{center}
\tablecomments{Parameters estimated from 
\url{https://www.predsci.com/hmi/data_access.php}. $S$ is the classical Lundquist number, 
$S^{\prime}$ gives the range expected from Langmuir waves to lower-hybrid waves, with the Coulomb collision frequency in the classical
value replaced by the wave frequency.}
\end{table*}

We model the electron distribution function as a $\kappa$-function, which offers a means of interpolating
between a quasi-Maxwellian core distribution and a power law tail at high velocities. We express the 
$\kappa$-function in terms of Maxwellians as follows,
\begin{eqnarray}
\nonumber f_{\kappa}\left(v\right)&=&\left[1+v^2/2\kappa v_t^2\right]^{-\kappa}\\
\nonumber &\simeq& \big[\exp\left(-v^2/2v_t^2\right)+\exp\left(-v^2\left(1+5/\kappa ^3\right)/6\right)/2\kappa\\
\nonumber & & +\exp\left(-v^2\left(1+5/\kappa ^3\right)/18\right)\times4/\kappa ^4\big]\\
& & /\left[1+1/2\kappa +4/\kappa ^4\right]
\end{eqnarray}
so that an atomic reaction rate for given $\kappa$ and temperature, $T$ is given by
\begin{eqnarray}
    \nonumber\sigma\left(\kappa, T\right)&=&\frac{\Gamma\left(\kappa\right)}{\Gamma\left(\kappa -3/2\right)\kappa ^{3/2}}\frac{1}{1+2/\kappa +4/\kappa ^4} \\\nonumber \times
    \bigg\{\sigma _{\rm M}\left(T\right)&+&\sqrt{\frac{3}{1+5/\kappa ^3}}\frac{\sigma _M\left(3T/\left(1+5/\kappa ^3\right)\right)}{2\kappa}\\&+&\sqrt{\frac{9}{1+5/\kappa ^3}}\frac{\sigma _M\left(9T/\left(1+5/\kappa ^3\right)\right)}{4\kappa ^4}\bigg\}
\end{eqnarray}
where $\Gamma\left(\kappa\right)$ is the Gamma function and $\sigma _M\left(T\right)$ is the Maxwellian rate coefficient evaluated at temperature $T$. Such an expansion is useful, because it allows ionization and recombination rates
appropriate for a $\kappa$ distribution to be constructed from rates readily available in tabulations
calculated by integrating the cross sections over a Maxwellian. \citet{hahn15} give a considerably more detailed expansion of the $\kappa$ distribution in terms of Maxwellians, but use a slightly different definition of $\kappa$. Our approximations for the $\kappa$ functions (thin sold lines) are compared in Fig. 2 with the true $\kappa$ function (thick solid lines) and the underlying Maxwellian (dashed lines), where $x=v/v_t$, and consider that our simpler approach is clearly adequate for the task at hand. With the identification of $\kappa = \gamma /2$ in equation 9, the values of $\kappa$ (or $\gamma$) can be interpreted in terms of reconnection parameters.

The energy of the $\kappa$-distribution is divided with proportion $5/2\kappa$ in the high energy tail and the remainder in the quasi-thermal
bulk. As $\kappa \rightarrow 5/2$ the energy of the distribution becomes infinite. This is resolved by invoking a cut off at high energies. In fact 
such a cut off is necessary because otherwise energy would accumulate in electrons with velocities greater than the speed of light, $c$, a clearly unphysical
situation. We implement this by setting a minimum value of $\kappa = 3$ when partitioning the energy between bulk and tail portions of the
distribution function. For $\kappa = 2.5$ and 2 this implies high velocity cutoffs at 27.4 and 14.2 times the thermal velocity of the bulk electrons, 
respectively, which for a bulk electron temperature of $\sim 10^6$ K, mean $1\times 10^{10}$ and $5\times 10^9$ cm s$^{-1}$ respectively,
i.e. suitably below $c$, and consistent with simulations in \citet{arnold21}. This is important in evaluating the collisional relaxation of the electron
distribution back to a Maxwellian, i.e. the heating of the ``bulk'' plasma electrons by the suprathermal tail, but less so in calculating ionization and recombination rates because the high energy electrons have smaller cross sections for these processes. We assume that the electron $\kappa$-distribution holds while the plasma is being heated by the reconnection, followed by relaxation back to a Maxwellian when this heating ceases. Within our model this happens when the CME acceleration ceases. In terms of the other timescales connected with the plasma evolution, this relaxation back to a  Maxwellian is essentially ``instantaneous''.

\begin{figure*}[t]
\centerline{\includegraphics[width=7.0in]{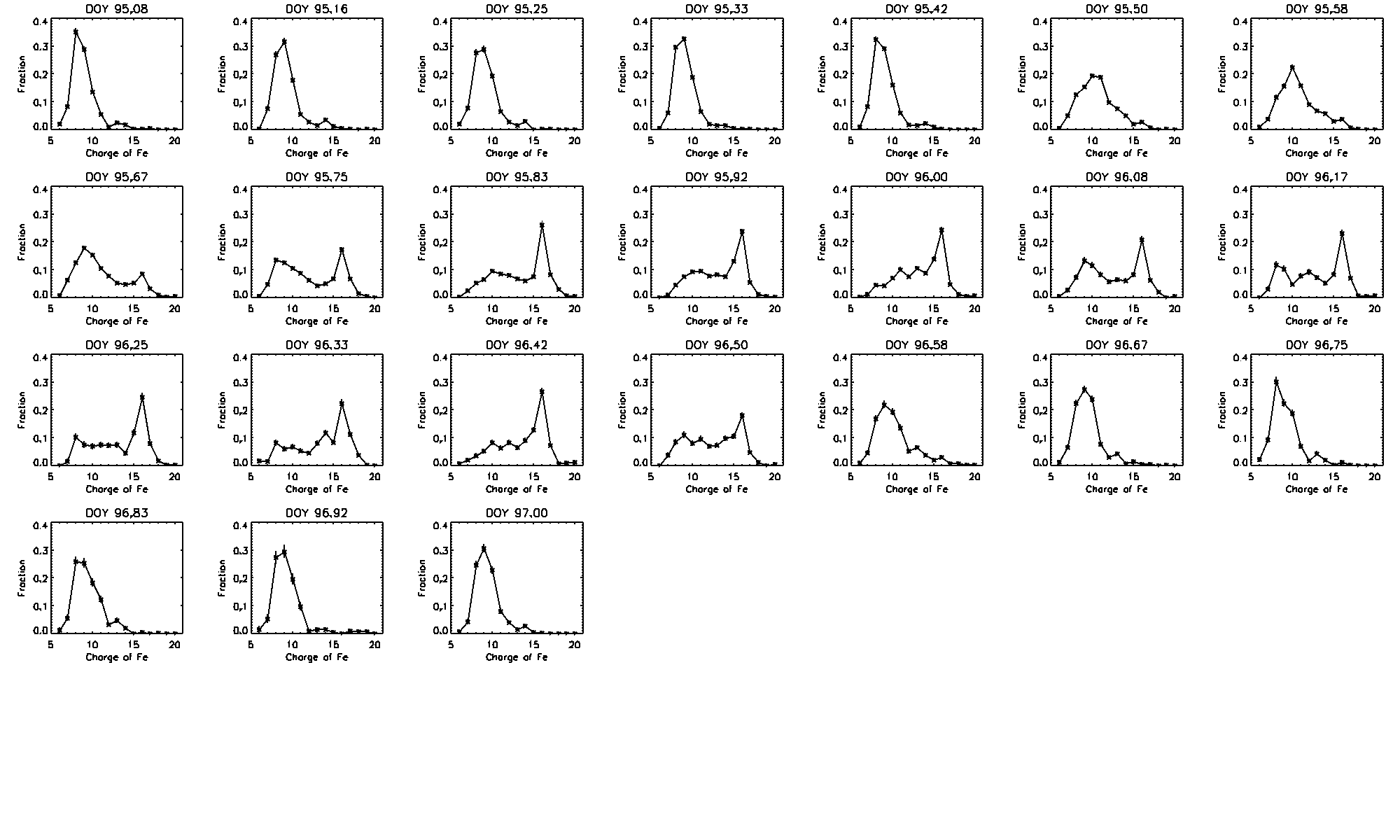}}
\vspace{-0.75true in}
\caption{Plots of the Fe charge state distribution for Day of Year (DOY) 95.08 - 97.00 during the 2010 April 3 ICME, observed at ACE during 2010 April 5-7.
During DOY 95.83 - 96.42, the dominant Fe charge state is 16+, indicating extra heating and ionization beyond the typical solar wind value of 8+ - 9+. \label{Festate}}
\end{figure*}

\section{Observations}
\subsection{Introduction}
In selecting the useful events our first criteria were observations of high charge states within the ICME, a CME eruption site
observed in the solar disk, and a likely post-CME current sheet with outflowing blobs. Accordingly, we looked for events
when the Solar TErrestrial RElations Observatory (STEREO) was in near quadrature with Earth, so that the STEREO-Sun-Earth angle 
was between 65 and 115 degrees.
This corresponds to dates between 2010 February 1 and 2012 March 1, a period which is further restricted by 
Advanced Composition Explorer/Solar Wind Ion Composition Spectrometer (ACE/SWICS)
data availability after August 2011. Therefore the promising events appear to be those on 2010 April 3, 2010 May 23, 2011 February 15,
2011 May 25 and 2011 June 2.

\subsection{2010 April 3 Event}
We concentrate first on the 2010 April 3 event. The eruption was observed by the STEREO Extreme Ultraviolet Imager (EUVI) of the Sun Earth Connection Coronal and Heliospheric Investigation \citep[SECCHI;][]{howard08} and the Extreme ultraviolet Imaging Telescope (EIT) on the Solar and Heliospheric Observatory (SOHO) by \citet{liu11}, by \citet{xie12} who also use WIND/WAVES radio data and by \citet{seaton11} who also include EUV observations from the Sun Watcher with Active Pixels and Image Processing 
\citep[SWAP;][]{berghmans06,degroof08} on the PROBA2 spacecraft. The CME is followed out further into the heliosphere using white light images
from SECCHI by \citet{liu11}, \citet{mishra20}, and \citet{xie12}. We use these works to define the expansion of the flux rope containing the CME ejecta, which is seen to erupt with a speed of about 64 km s$^{-1}$ \citep{seaton11}, starting from a heliocentric distance of 
1.2 $R_{\sun}$ \citep{seaton11,zuccarello12}, with electron density $2\times 10^9$ cm$^{-3}$, beginning at about 8:50 UT on 2010 April 3. This is actually a few minutes before the onset of the flare
observed by GOES. \citet{mishra20} pick up the CME and flux rope expansion from about 2.5 $R_{\sun}$. The flux rope continues to accelerate at 0.05 km s$^{-2}$ and
expand until a speed of about 700 km s$^{-1}$ is reached at a heliocentric radius of about 10 $R_{\sun}$. Other authors give the maximum speed as 
800 km s$^{-1}$ \citep{liu11}, $\sim 900$ km s$^{-1}$ \citep{mostl10, wood2011}, or $\sim 1000$ km s$^{-1}$ \citep{xie12}. The ICME at Earth orbit is observed
with speeds 550 - 700 km s$^{-1}$ \citep{Rouillard2011,liu11,mostl10,rodari19,xie12}.

We characterize coronal plasma into which the CME expands using simulation results provided by Predictive Sciences at 
\url{https://www.predsci.com/hmi/data_access.php}. The electron density and Alfv\'en speed at various heliocentric radii are given in Table 1, along
with various estimates of the Lundquist number. The first is the ``classical'' value, $S$, assuming an electrical conductivity given by its Spitzer value, $\sigma = \omega _{pe}^2/4\pi\nu _c$ where $\omega _{pe}$ is the electron plasma frequency and 
$\nu _c$ is the Coulomb collision frequency. The second range of $S^{\prime}$ is given by replacing $\nu _c$ in the Spitzer value by the local electron plasma or lower-hybrid frequencies.

\begin{table*}[t]
\begin{center}
\caption{2011 February 15 CME Off-Limb Plasma Parameters}
\begin{tabular}{cccccc}
\hline Radius ($R_{\sun}$) & $V_A$ (km s$^{-1}$)  & $\int V_Adl$ & $n_e$& $S=\int V_Adl\sigma/4\pi c^2$ & $S^{\prime}$ \\\hline
1.1 & 1500 & 150 &$6\times 10^8$  & $9\times 10^{11}$ & $3\times 10^4 - 2\times 10^8$ \\
1.5 & 750& 450  &$8\times 10^6$  & $2\times 10^{12}$&  $6\times 10^4 - 4\times 10^8$\\
2.0 & 500& 700 & $2\times 10^6$ & $4\times 10^{12}$& $ 4\times 10^4 - 3\times 10^8$\\
2.5 & 500& 950& $7\times 10^5$ & $5\times 10^{12}$& $3\times 10^4 - 2\times 10^8$\\
3.0 & 500& 1200& $4\times 10^5$ & $6\times 10^{12}$& $2\times 10^4 - 2\times 10^8$\\
5.0 & 300& 1800& $1\times 10^5$ & $1\times 10^{13}$&  $1.3\times 10^4 - 1\times 10^8$\\
\hline  \end{tabular}
\end{center}
\tablecomments{Parameters estimated from \url{https://www.predsci.com/hmi/data_access.php}. $S$ is the classical Lundquist number, 
$S^{\prime}$ gives the range expected from Langmuir waves to lower-hybrid waves, with the Coulomb collision frequency in the classical
value replaced by the wave frequency. }
\end{table*}

The ICME related to this eruption is studied in-situ using data from the Advanced Composition Explorer (ACE), WIND, SOHO and STEREO
\citep{rodari19,mostl10,liu11}. We focus on the ACE/SWICS charge state observations. In Fig. \ref{Festate} we show 24 epochs  (i.e. integrated 2 hour observations) 
of Fe charge state distributions
recorded by ACE/SWICS. The first panel shows data from Day Of Year (DOY) 95.08, April 5 01:55 UT, and is dominated by slow speed solar wind where Fe 
charge states 8+ - 9+ are the most abundant. At DOY 95.83 - 96.42 (panels 9 - 18) the dominant Fe charge state is now 16+, presumably as a result of heating and ionization in the reconnection current sheet as the CME erupts. The later panels show a return to Fe charge states around 8+ - 9+, typical of the 
slow speed solar wind. A concentration of Fe in the 16+ charge state can be suggestive of strong heating followed by recombination \citep{gruesbeck11}. 
This Ne-like charge state has weak recombination rates into the next lower charge state due to the closed shell structures,
and so in recombining conditions, Fe charges bottleneck here. However for this event (c.f. the 2011 February 15 event below), although Fe can
be explained in this way, the other elements, C, O, Mg, and Si do not follow the same trend, being heavily overionized. In fact modeling these five elements
simultaneously is highly constraining, and is important to our eventual conclusions.

\begin{table*}[t]
\begin{center}
\caption{Model Fits 2010/04/03 CME}\label{aprcme}
\begin{tabular}{ccccccccccccc}
\hline
& & \multicolumn{11}{c}{$\kappa$}\\\hline
epoch& year DOY time & $\infty $ & $10^4$ & $10^2$& 10 & 4.5 & 4 & 3.5& 3.1&  2.6& 2.0& 1.6\\ \hline
10& 2010 95 19:58 heat& 0.31& 0.26& 0.26& 0.32& 0.50& 0.60& 0.80& 1.25& 1.50& 1.70& 8.0\\
& $\Sigma\Delta f_q^2$& 0.38& 0.40& 0.39& 0.38& 0.37& 0.36& 0.35& 0.35& 0.34& 0.33& 0.35\\ &&&&&&&&&&&&\\  
11& 2010 95 21:58 heat& 0.21& 0.23& 0.24& 0.29& 0.45& 0.55& 0.70& 1.15& 1.30& 1.45& 6.2\\
& $\Sigma\Delta f_q^2$& 0.26& 0.27& 0.27& 0.26& 0.23& 0.23& 0.22& 0.21& 0.20& 0.21& 0.23\\ &&&&&&&&&&&&\\
12& 2010 95 23:58 heat&  0.21& 0.23& 0.23& 0.28& 0.45& 0.55& 0.70& 1.10& 1.25& 1.45& 6.2\\
& $\Sigma\Delta f_q^2$& 0.33& 0.30& 0.29& 0.29& 0.29& 0.29& 0.28& 0.27& 0.26& 0.26& 0.26\\ &&&&&&&&&&&&\\
13& 2010 96 01:58 heat& 0.24& 0.26& 0.26& 0.32& 0.50& 0.60& 0.80& 1.25& 1.50& 1.70& 8.6\\
& $\Sigma\Delta f_q^2$& 0.32& 0.31& 0.31& 0.30& 0.29& 0.29& 0.28& 0.27& 0.26& 0.27& 0.27\\ &&&&&&&&&&&&\\
14& 2010 96 03:58 heat& 0.29& 0.32& 0.32& 0.40& 0.65& 0.75& 1.05& 1.65& 1.85& 2.30& 13.\\
& $\Sigma\Delta f_q^2$& 0.34& 0.34& 0.34& 0.33& 0.32& 0.32& 0.31& 0.30& 0.29& 0.29& 0.29\\&&&&&&&&&&&& \\
15& 2010 96 05:58 heat& 0.26& 0.28& 0.28& 0.34& 0.55& 0.65& 0.85& 1.25& 1.55& 1.85& 9.4\\
& $\Sigma\Delta f_q^2$& 0.28& 0.28& 0.28& 0.27& 0.26& 0.26& 0.25& 0.25& 0.23& 0.23& 0.24\\&&&&&&&&&&&& \\
16& 2010 96 07:58 heat& 0.27& 0.29& 0.30& 0.36& 0.55& 0.65& 0.90& 1.45& 1.65& 1.95& 10.4\\
& $\Sigma\Delta f_q^2$& 0.30& 0.30& 0.30& 0.29& 0.27& 0.26& 0.25& 0.24& 0.22& 0.23& 0.24\\ &&&&&&&&&&&&\\
17& 2010 96 09:58 heat& 0.23& 0.24& 0.24& 0.30& 0.50& 0.55& 0.75& 1.20& 1.35& 1.55& 6.9\\
& $\Sigma\Delta f_q^2$& 0.30& 0.33& 0.33& 0.30& 0.27& 0.26& 0.24& 0.24& 0.22& 0.23& 0.26\\ &&&&&&&&&&&&\\
18& 2010 96 11:58 heat& 0.22& 0.24& 0.24& 0.29& 0.45& 0.55& 0.75& 1.15& 1.30& 1.50& 6.4\\
& $\Sigma\Delta f_q^2$& 0.42& 0.45& 0.45& 0.42& 0.39& 0.37& 0.36& 0.35& 0.34& 0.35& 0.39\\\hline
\end{tabular}
\end{center}
\end{table*}

\subsection{2011 February 15 Event}
The 2011 February 15 CME was observed in white light by both STEREO satellites and SOHO/LASCO \citep{gopalswamy12} and additionally by 
Wind/WAVES \citep{jing15}. With the flare starting at 01:46:50 UT, and peaking at 01:54:08 UT \citep{gopalswamy12}, the CME is observed to
appear in the LASCO C2 coronagraph (at 2.2 R$_{\sun}$ heliocentric distance) at 02:24 UT, already traveling at its final speed of $\sim 1000$ km s$^{-1}$.
The average speed between the solar surface and 2.2 R$_{\sun}$ is 500 km s$^{-1}$ and the minimum acceleration is 0.5 km s$^{-2}$. We find
the best matches to the observed charge states for an initial acceleration of 1.4 km s$^{-2}$, going to zero when the speed reaches 1000 km s$^{-1}$.
This acceleration matches well with that given by \citet{maricic14}, from analysis of STEREO/SECCHI white light and EUV data. These authors analyze
three CMEs erupting form the same active region, and \citet{temmer14} investigate the interaction between the 2011 February 14 and 15 events in more 
detail. 

We take an initial height for the CME flux rope of 0.07 R$_{\sun}$ above the solar surface, (1.07 R$_{\sun}$ heliocentric distance), \citep{cho13,jing15}, and an initial
velocity of 75 km s$^{-1}$ from \citet{kay17}. The initial density is $5\times 10^9$ cm$^{-3}$, specified to match that measured at  10.5 R$_{\sun}$ \citep{temmer14}. This also matches the density measured at 1 AU, after expanding according to equation 8 with 
$r_{\rm Alf}=10$ R$_{\sun}$. The off-limb electron density and Alfv\'en speed with estimate of the Lundquist number are given in Table 2 similarly for the 2010 April 3 event.

\begin{figure*}[t]
\centerline{\includegraphics[width=3.5in]{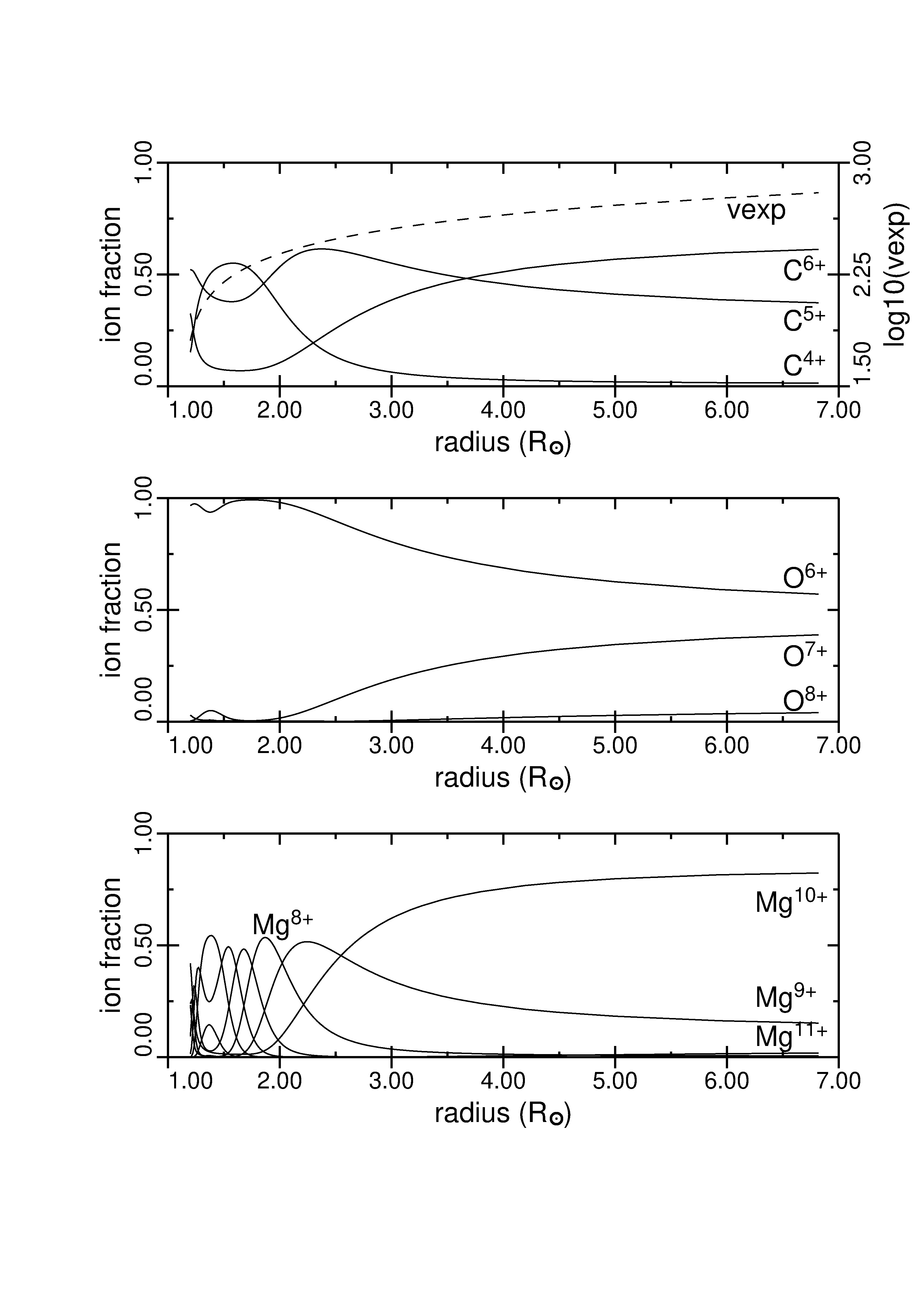}
\includegraphics[width=3.5in]{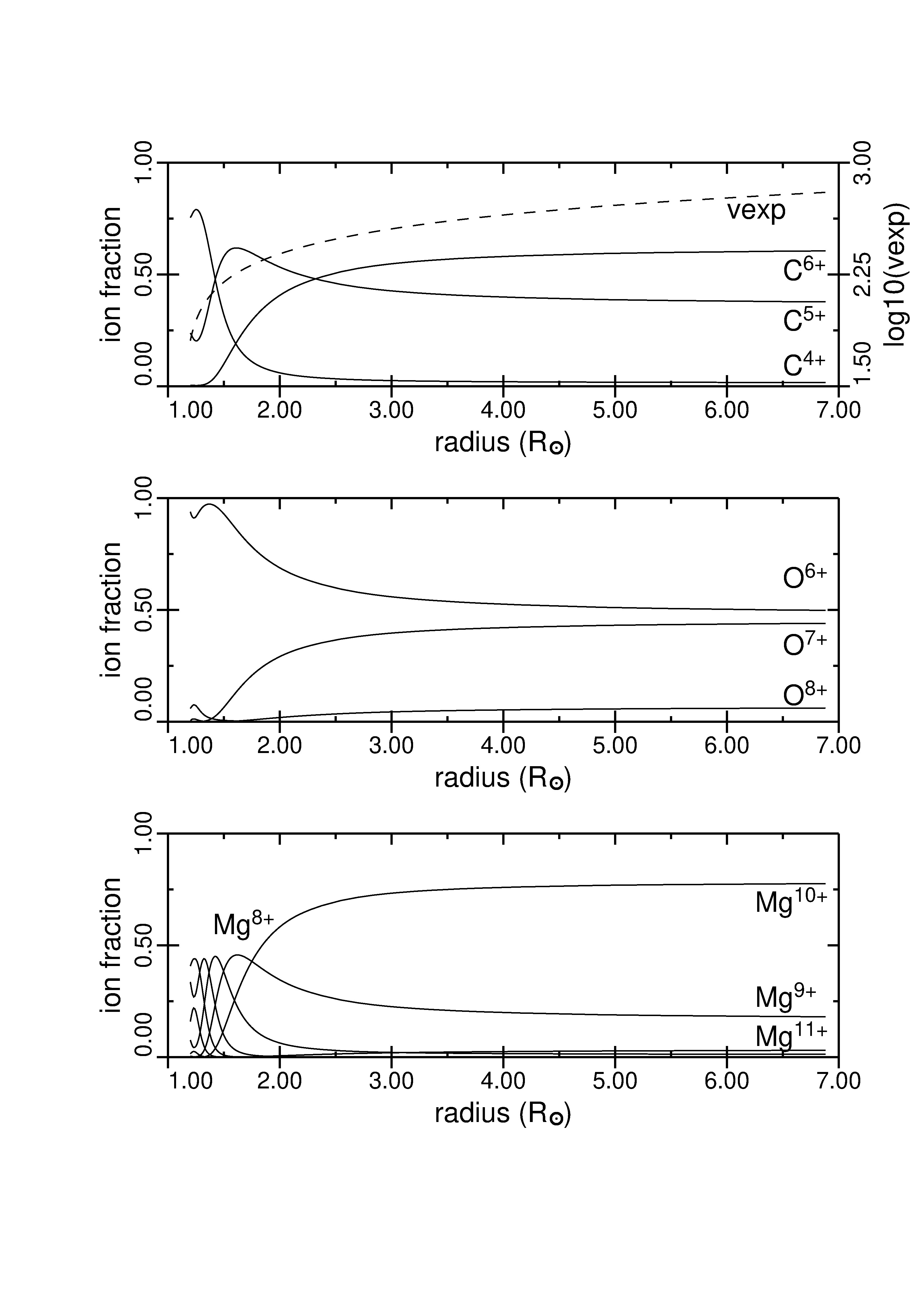}}
\vspace{-0.5true in}
\centerline{\includegraphics[width=3.5in]{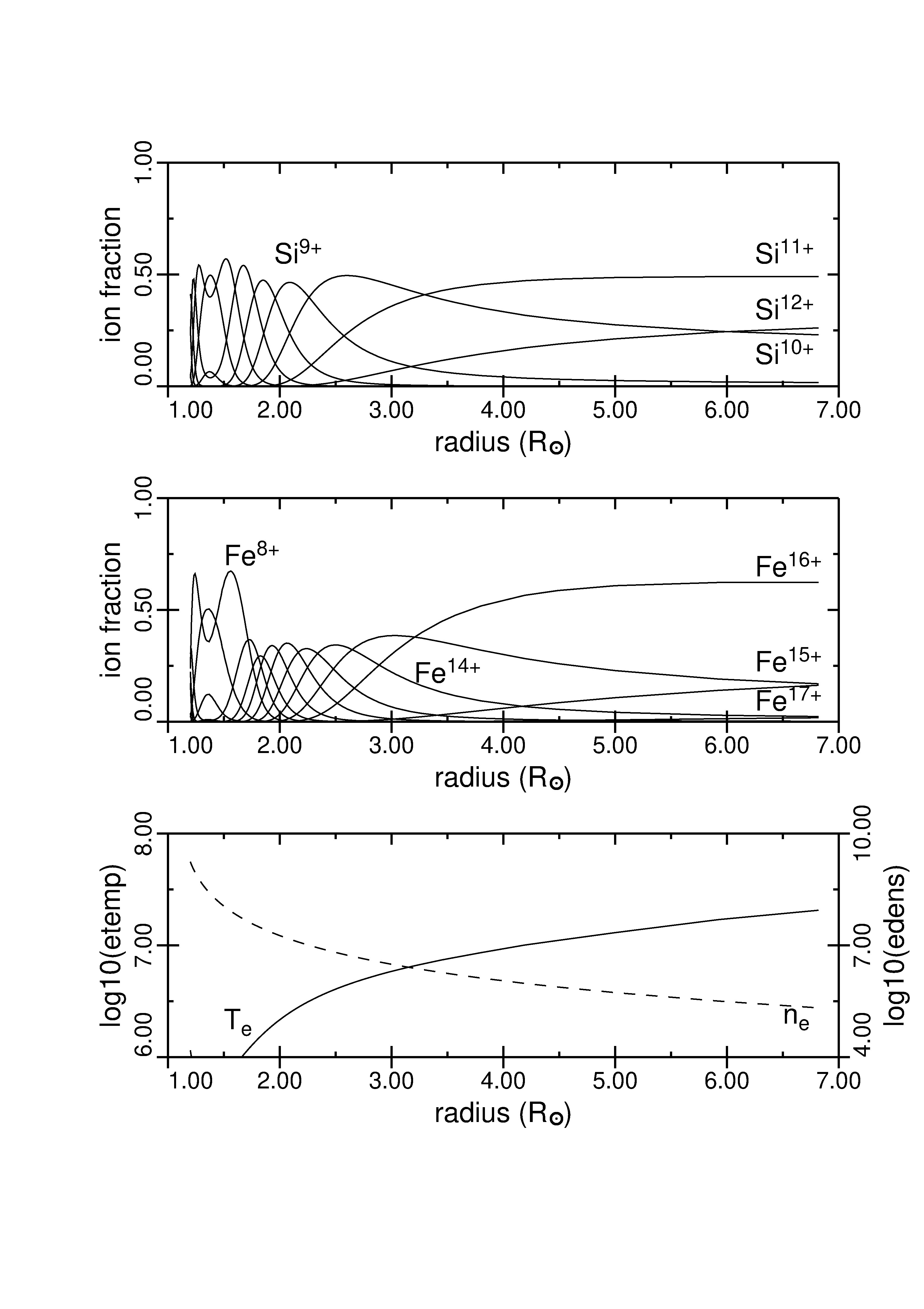}
\includegraphics[width=3.5in]{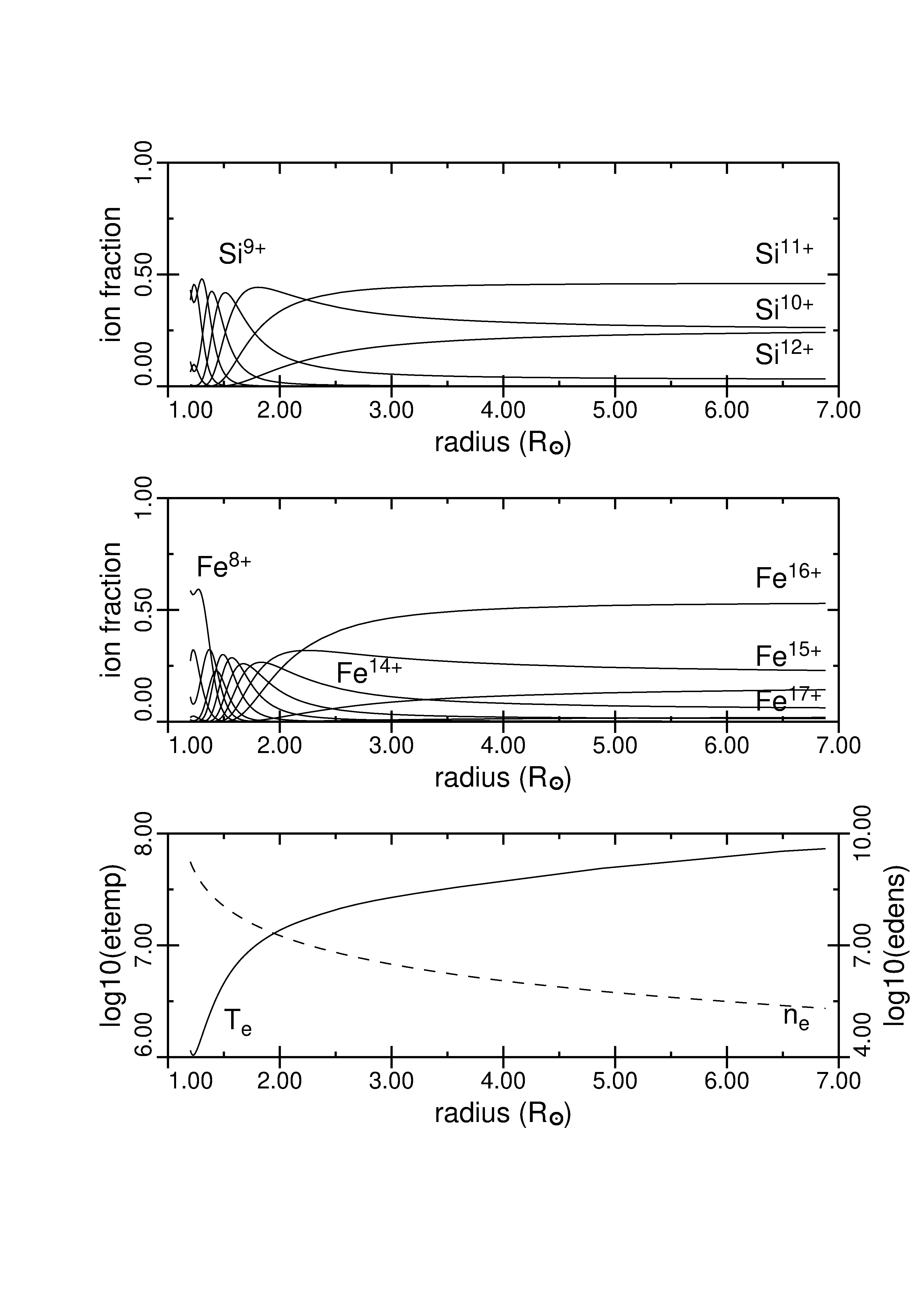}}
\vspace{-0.75true in}
\caption{Variation of ionization balance with heliocentric distance for, from top C, O, Mg, Si and Fe, with electron temperature and density profiles given at the bottom, for the 2010 April 03
CME ejecta epoch 16 (from Table 3). The left panels give the case for Maxwellian electron distributions. The right panels are the same but for $\kappa = 2.6$ electron distributions. The broader charge state distributions are clearest for Si and Fe. The CME expansion velocity is superimposed on the top panel, to be read on the right hand axis. Selected charge states are labeled. Others can be inferred by comparsion with Fig. 5.
\label{hifibo}}
\end{figure*}

\begin{figure*}[t]
\centerline{\includegraphics[width=3.5in]{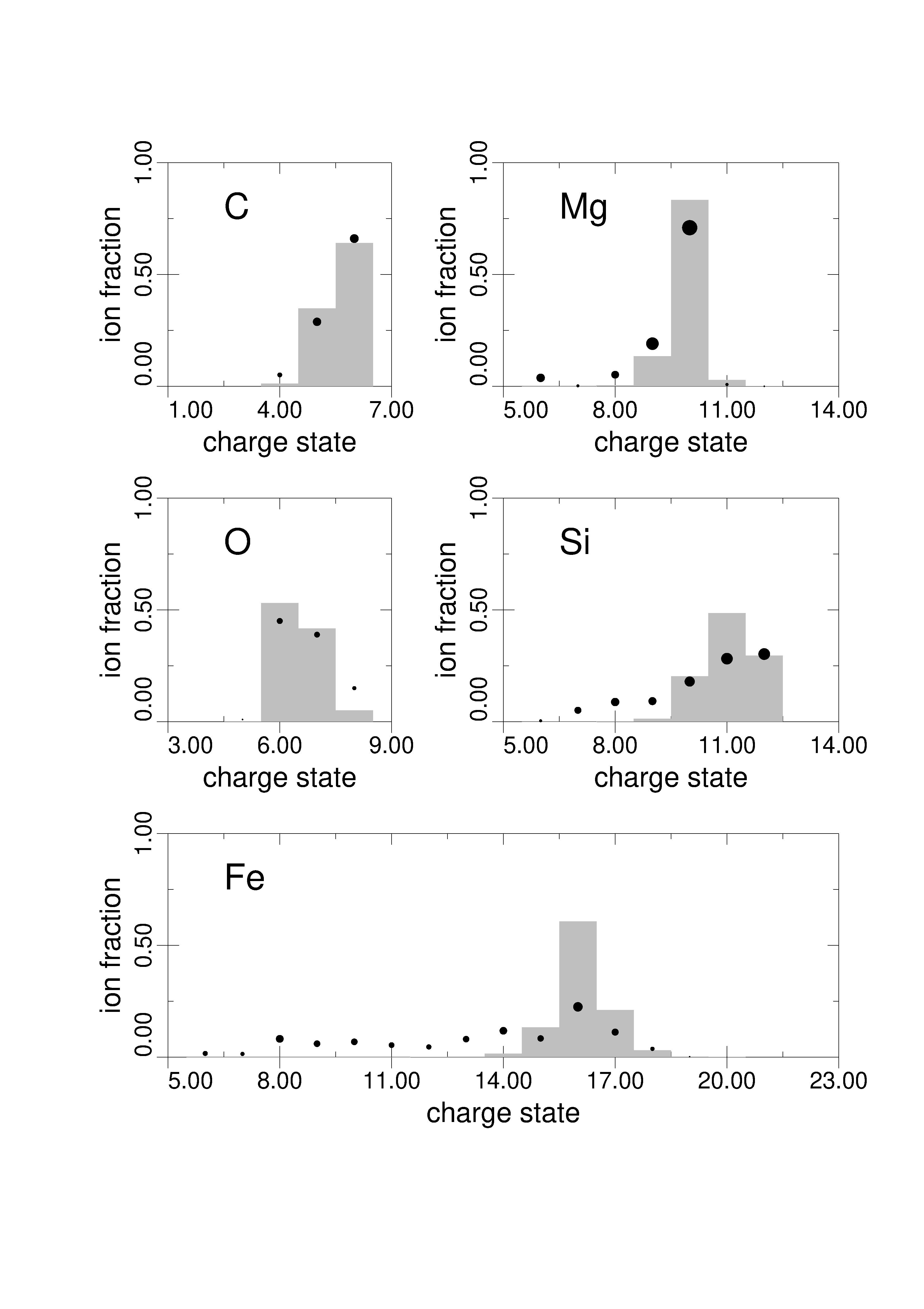}
\includegraphics[width=3.5in]{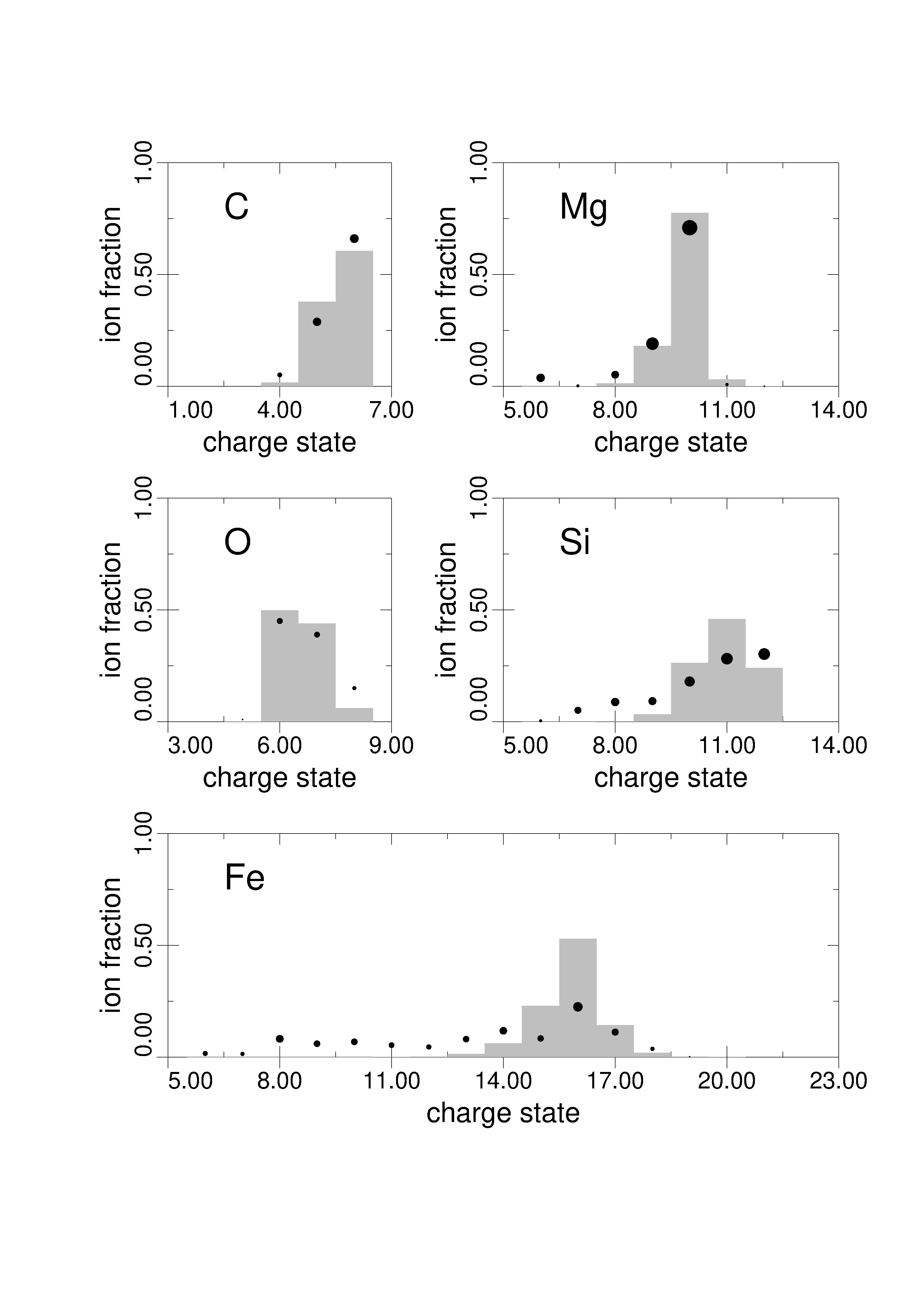}}
\vspace{-0.5true in}
\caption{Left: Final charge state distributions for C, O, Mg, Si and Fe, at 1 AU in epoch 16 in ICME ejecta from the 2010 April 03 CME event assuming Maxwellian electron distributions. The model is shown as a histogram, and the data are shown as filled
black circles, with 2$\sigma$ uncertainties from the level 2 ACE data given by the radius. Right: The same but for $\kappa = 2.6$ electron distributions. The difference in modeled charge states distribution between two cases is most clearly seen for Fe and Si.
\label{hificmecs}}
\end{figure*}

For the X2.2 flare associated with the 2011 February 15 CME, we have estimated the photoionization rate of Fe$^{15+}\rightarrow $ Fe$^{16+}$ to compare with collisional ionization. Using various power laws or thermal bremsstrahlung for the radiation spectrum, we find the ratio of rates for photoionization/collisional ionization to be about 0.25\% at 2 $R_{\sun}$ (assumed flare location at 1 $R_{\sun}$), and a maximum of about 10\% at 6 $R_{\sun}$. Given that the charge states freeze in quickly for this event, this is not high enough to concern us (errors on atomic data are typically or order 10\%, so we neglect this extra complication here. Clearly, for events like the Halloween flare of 2003 at X28 - X45, photoionization will be more important. The 2010 April 3 event had a much less luminous flare, at B7.4, so we do not consider photoionization there at all.

\subsection{Other Events}
The 2010 April 3 and 2011 February 15 are the best observed events for our purposes. The 2010 May 23 CME is studied by \citet{lugaz12}, who concentrate on its interaction with a following CME on 2010
May 24. \citet{song12} identify merging magnetic islands and an ensuing Type III radio burst which 
is an indicator of electron acceleration, precisely the mechanism under study in our paper. However this particular merger occurred at heliocentric radii greater than $4R_{\sun}$, just beyond the region where charge states are typically frozen-in, and so is unlikely to affect the evolution of the ionization balance of the CME ejecta. For the 2011 May 25 and 2011 June 2 CMEs, we are unaware of 
analysis of data early in the CME evolution, while they are still accelerating. CME height-time plots
beginning at about 3 $R_{\sun}$ are given by \citet{gopalswamy19} for the 2011 June 2 event, but their main interest is in the shock wave, Type II radio bursts and the sustained gamma ray emission,
while \citet{palmerio17} are more focused on the magnetic structure of the flux rope.

\section{Models}
\subsection{2010 April 3 Event}
We model the charge state distributions for C, O, Mg, Si, and Fe at each epoch by integrating Equations \ref{denseq}, \ref{iontemp}, and \ref{etemp} within background hydrodynamic 
expansion taken from the observations cited above. A Lagrangian plasma parcel starts from 1.2 $R_{\sun}$ heliocentric distance with initial speed of
64 km s$^{-1}$, and an acceleration of 0.05 km s$^{-1}$ leading to a final speed of 1000 km s$^{-1}$. The initial electron density is taken to be 
$2\times 10^9$ cm$^{-3}$, which when expanded to 1 AU from 6.5 $R_{\sun}$ by $1/r^3$ gives the observed {\it in situ} density of $\sim 5-10$ cm$^{-3}$ \citep{liu11,mostl10,xie12,rodari19}. The initial temperature is typical of the corona, $1.15\times 10^6$K. 
These parameters mean that the CME acceleration, and by assumption the heating of its ejecta, extend to 
radii well beyond that where the charge states freeze-in. Thus the entire charge state fraction evolution takes place in a non-Maxwellian electron
distribution function.

The heating of electrons is
assumed to be a fraction of the gain in kinetic and gravitational energy of the CME ejecta. We vary model parameters until quantity
$\Sigma\Delta f_q^2 = \Sigma \left(f_{q,obs}-f_{q,mod}\right)^2$ is minimized. This is the sum of differences between observed and modeled
charge state fractions squared, taken over all ions of all five elements considered. A number of other choices might be possible here, but we adopt
this prescription for its simplicity, and for its relative weighting towards the more populated charge state fractions observed. One could weight the
sum of differences squared by the observational errors, or weight each element rather than each observed charge state equally. We point out here
that the observed charge state fractions we are dealing with are not independent variables, since the charge state fractions for any element must add 
up to unity. Therefore the usual statistical approaches and theorems regarding least squares minimization will not be valid. Rather than rely on such
techniques, we analyze a sample of charge state observations from this and other CMEs with different evolutionary histories to understand how results
are distributed.

\begin{table*}[t]
\begin{center}
\caption{Model Fits 2011/02/15 CME}\label{febcme}
\begin{tabular}{ccccccccccccc}
\hline
& & \multicolumn{11}{c}{$\kappa$}\\\hline
epoch& year DOY time & $10^6$ & $10^4$ & $10^2$& 10 & 4.5 & 4 & 3.5& 3.1&  2.6& 2.0& 1.6\\ \hline
9& 2011 49 16:43 heat&   0.12& 0.12& 0.12& 0.12& 0.15& 0.17& 0.22& 0.40& 0.35& 0.10&0.11\\ 
& $\Sigma\Delta f_q^2$& 0.29& 0.29& 0.28& 0.25& 0.19& 0.17& 0.15& 0.14& 0.15& 0.21& 0.49\\  &&&&&&&&&&& \\
10& 2011 49 18:43 heat& 0.11&0.11 &0.11 & 0.12& 0.15& 0.17& 0.22& 0.35& 0.29& 0.11& 0.11\\
& $\Sigma\Delta f_q^2$& 0.45&0.45 &0.45 & 0.40& 0.30& 0.27& 0.21& 0.16& 0.18& 0.32& 0.64\\ 
&&&&&&&&&&& \\
11& 2011 49 20:44 heat& 0.10& 0.10& 0.103& 0.10& 0.13& 0.15& 0.19& 0.35& 0.29& 0.10& 0.11\\
& $\Sigma\Delta f_q^2$& 0.42& 0.43& 0.41& 0.38& 0.31& 0.28& 0.24& 0.21& 0.22& 0.29& 0.67\\  
&&&&&&&&&&& \\
12& 2011 49 22:44 heat& 0.11 & 0.11& 0.11& 0.12 & 0.14& 0.15& 0.18& 0.60& 0.23& 0.12& 0.11\\
&$\Sigma\Delta f_q^2$&0.42 & 0.42& 0.41& 0.37& 0.31& 0.30& 0.29& 0.30& 0.31& 0.30& 0.54\\  
&&&&&&&&&&& \\
15& 2011 50 04:44 heat& 0.10& 0.10& 0.10& 0.10& 0.13& 0.14& 0.15& 0.25& 0.22& 0.10& 0.11\\
& $\Sigma\Delta f_q^2$& 0.39& 0.39& 0.39& 0.35& 0.30& 0.27& 0.23& 0.17& 0.18& 0.32& 0.82\\  
&&&&&&&&&&& \\
18& 2011 50 10:44 heat& 0.11& 0.11& 0.11& 0.12& 0.14& 0.15& 0.15& 0.35& 0.30& 0.12& 0.11\\
& $\Sigma\Delta f_q^2$& 0.28& 0.28& 0.28& 0.25& 0.21& 0.20& 0.18& 0.23& 0.22& 0.22& 0.49\\  &&&&&&&&&&& \\
19& 2011 50 12:45 heat& 0.12& 0.12& 0.12& 0.13& 0.14& 0.16& 0.20& 0.35& 0.35& 0.12& 0.11\\  
& $\Sigma\Delta f_q^2$& 0.14& 0.14& 0.14& 0.15& 0.17& 0.19& 0.25& 0.32& 0.29& 0.19& 0.39\\  &&&&&&&&&&& \\
20& 2011 50 14:45 heat& 0.12& 0.12& 0.12& 0.13& 0.14& 0.18& 0.20& 0.40& 0.35& 0.12& 0.11\\
& $\Sigma\Delta f_q^2$& 0.35& 0.35& 0.34& 0.35& 0.41& 0.44& 0.49& 0.58& 0.38& 0.44& 0.62\\  &&&&&&&&&&& \\
21& 2011 50 16:45 heat& 0.12& 0.12& 0.12& 0.12& 0.14& 0.16& 0.20& 0.40& 0.35& 0.11& 0.11\\
&$\Sigma\Delta f_q^2$& 0.25& 0.25 & 0.25& 0.25& 0.26& 0.28& 0.33& 0.40& 0.38& 0.29& 0.49\\  &&&&&&&&&&& \\
22& 2011 50 18:45 heat& 0.13& 0.13& 0.13& 0.13& 0.16& 0.16& 0.20& 0.37& 0.45& 0.15& 0.11\\
& $\Sigma\Delta f_q^2$& 0.19& 0.19& 0.19& 0.18& 0.20& 0.19& 0.28& 0.39& 0.35& 0.20& 0.30\\  &&&&&&&&&&& \\
25& 2011 51 00:46 heat & 0.11& 0.11& 0.11& 0.11& 0.14& 0.15& 0.19& 0.30& 0.22& 0.11& 0.11\\
& $\Sigma\Delta f_q^2$& 0.22& 0.22& 0.21& 0.18& 0.12& 0.10& 0.09& 0.11& 0.10& 0.15& 0.48\\  &&&&&&&&&&&& \\
27& 2011 51 04:46 heat & 0.10& 0.10& 0.10& 0.10& 0.11& 0.12& 0.15& 0.25& 0.19& 0.12& 0.11\\
& $\Sigma\Delta f_q^2$& 0.29& 0.29& 0.30& 0.26& 0.25& 0.23& 0.20& 0.19& 0.19& 0.35& 0.85\\  
&&&&&&&&&&&& \\
28& 2011 51 06:46 heat& 0.10& 0.10& 0.10& 0.10& 0.13& 0.15& 0.15& 0.25& 0.22& 0.12& 0.11\\
& $\Sigma\Delta f_q^2$& 0.35& 0.35& 0.35& 0.32& 0.27& 0.24& 0.20& 0.15& 0.17& 0.35& 0.79\\  
&&&&&&&&&&&& \\
31& 2011 51 12:47 heat& 0.11 & 0.10& 0.10& 0.10& 0.13& 0.12& 0.15& 0.30& 0.22& 0.12& 0.11\\
& $\Sigma\Delta f_q^2$& 0.31& 0.31& 0.30& 0.30& 0.21& 0.24& 0.18& 0.19& 0.18&  0.23& 0.56\\&&&&&&&&&&&& \\
32& 2011 51 14:47 heat& 0.10& 0.10& 0.11& 0.10& 0.12& 0.12& 0.14& 0.25& 0.19& 0.12& 0.11\\
& $\Sigma\Delta f_q^2$& 0.29& 0.29& 0.41& 0.26& 0.21& 0.19& 0.17& 0.19& 0.19& 0.22 & 0.69\\

\hline
\end{tabular}
\end{center}
\end{table*}

Results for epochs 10 - 18 are given in Table \ref{aprcme}, for a variety of
assumed values of $\kappa$, ranging from $10^4$ (effectively a Maxwellian) down to 1.6, i.e. close to the lower limit given by Equation 8, as well as for a pure Maxwellian ($\kappa\rightarrow\infty $). These epochs comprise 
the periods of the ICME with significant Fe$^{16+}$ (see Figure \ref{Festate}) and for which reasonable fits to the data can be found.
In all cases the lowest permissible $\kappa$'s in the range 2 - 3 give the best model fit to the observed charge states. As $\kappa$ increases towards
infinity the fits deteriorate monatonically. The heating required as a fraction of the increase in kinetic and gravitational energy increases for decreasing $\kappa$, 
and is similar to that predicted by Equation \ref{enfrac} for $S_e\sim 10^4$ for $\kappa = 2 - 3$. The $\kappa$ distributions are apparently less efficient at ionizing the CME ejecta,
presumably because so many electrons are placed at energies well above the ionization thresholds where the ionization cross sections are
decreasing with increasing electron energy.

Figures \ref{hifibo} and \ref{hificmecs} illustrate these calculations for epoch 16. Figure \ref{hifibo} shows the evolution of the C, O, Mg, Si, and 
Fe charge state fractions as a function of
heliocentric radius, together with the evolution of the electron temperature and density. The set of panels on the left show the results with a Maxwellian electron distribution, while on the right the model with $\kappa = 2$ is shown. Figure \ref{hificmecs} shows the final model charge state fractions for C, O, Mg, Si, and Fe, as histograms, compared with {\it in situ} observations as black circles. The difference betwen a Maxwellian and $\kappa$ distributions
is most clearly seen in the Fe and Si charge state distributions in Fig. \ref{hificmecs}, and also in the charge state evolutions in Fig.  \ref{hifibo}. 
The $\kappa$ electron
distribution produces a wider range of charge states than the Maxwellian. This is understandable, and to be expected, since the $\kappa$ distribution
encompasses a wider range of electron energies than does the Maxwellian. Excess population in lower charge states than those modeled is visible for both
Si and Fe. We attribute this to mixing of the heated CME ejecta with either unheated ejecta or ambient solar wind. Such mixing would also enhance
lower charge states of C, O , and Mg, but is less obvious here because these charge states are already populated anyway.

\begin{figure*}[t]
\centerline{\includegraphics[width=3.5in]{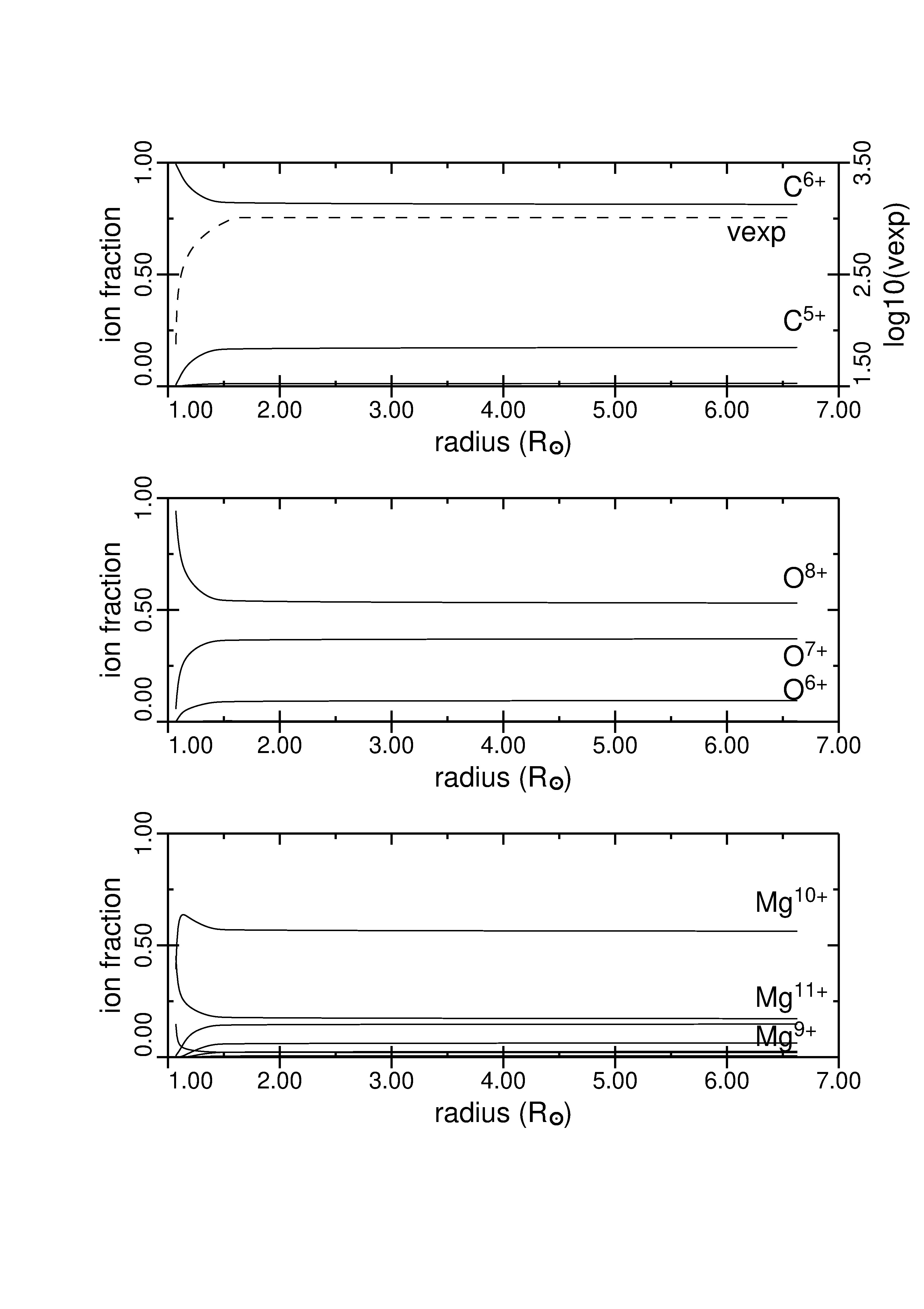}
\includegraphics[width=3.5in]{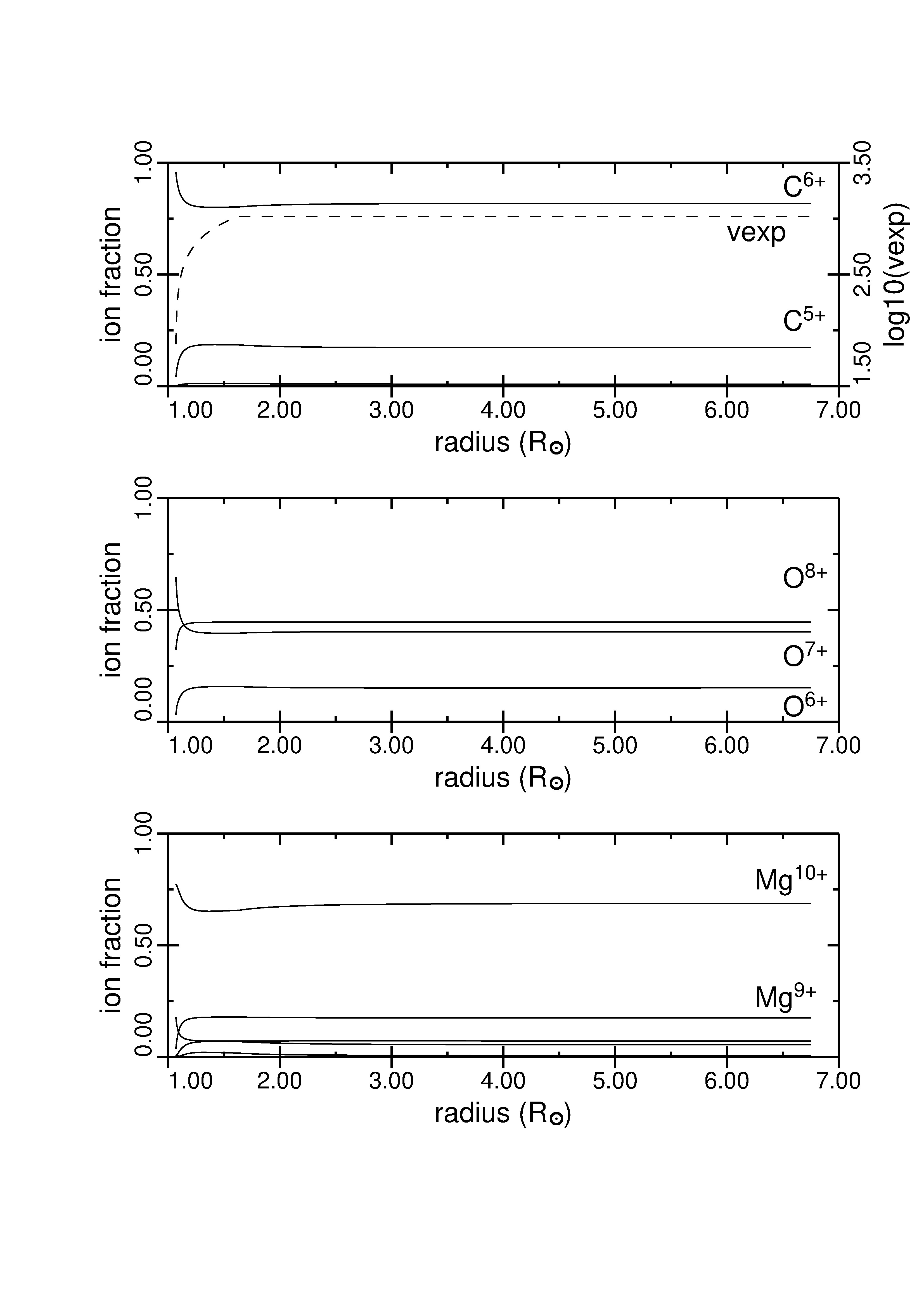}}
\vspace{-0.5true in}
\centerline{\includegraphics[width=3.5in]{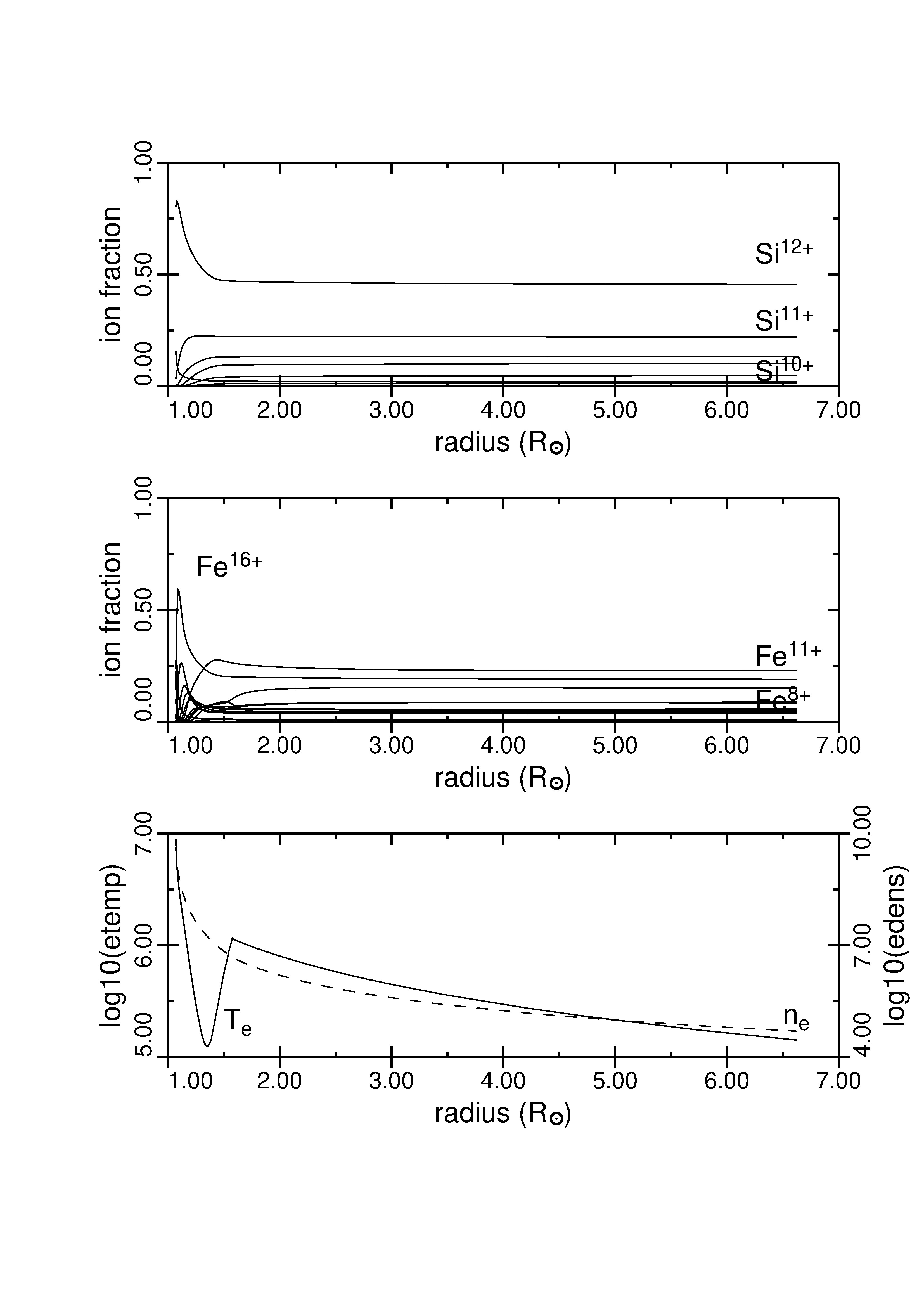}
\includegraphics[width=3.5in]{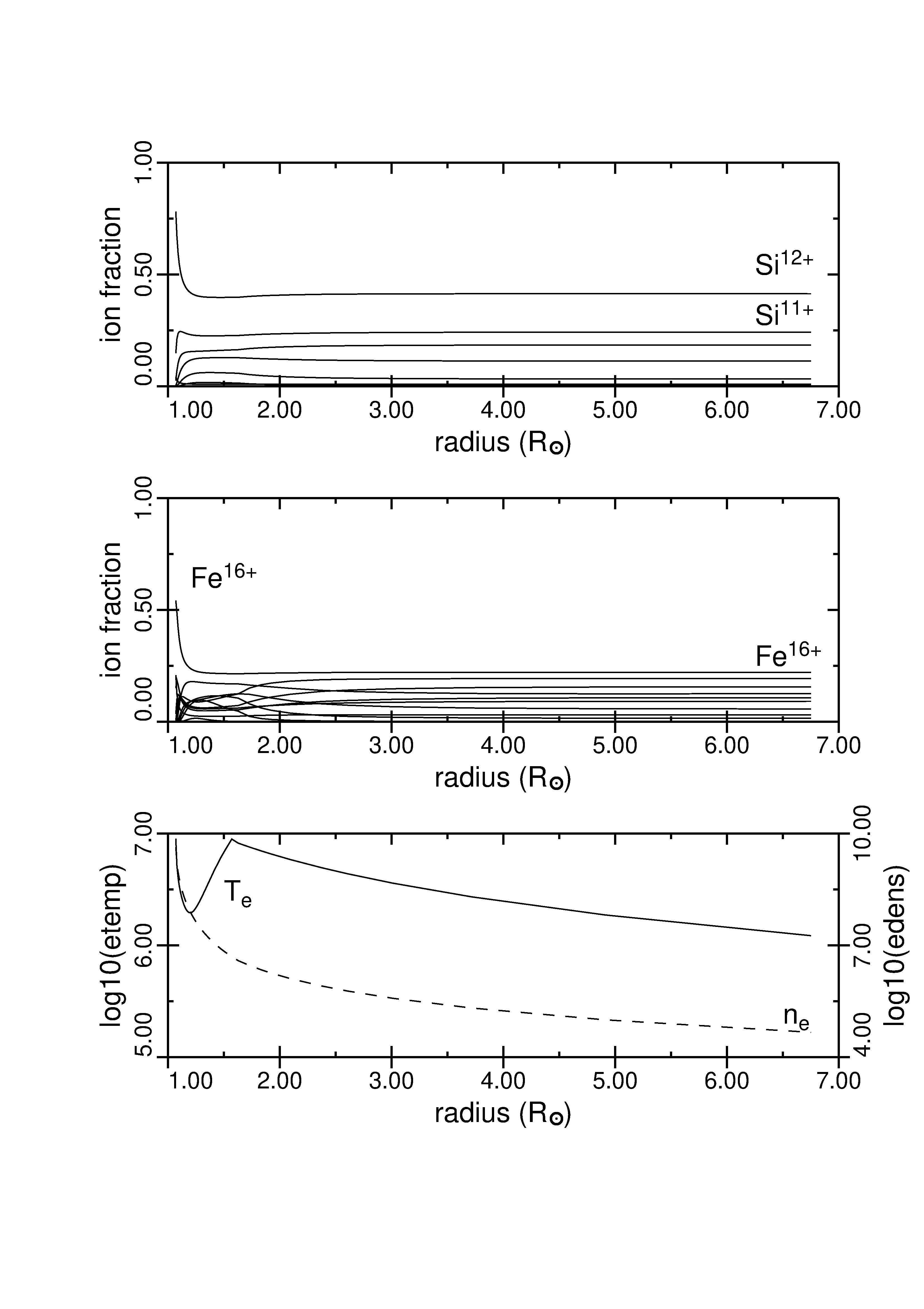}}
\vspace{-0.75true in}
\caption{Left: Variation of ionization balance with heliocentric distance for C, O, Mg, Si and Fe, and the electron temperature and density, for the 2011 February 15
CME ejecta epoch 25, assuming $\kappa = 10^6$ electron distributions. Right: The same but for $\kappa = 3.5$ electron distributions. The CME expansion velocity is superimposed on the top panel, to be read on the right hand axis. Selected charge states are labeled. Others can be inferred by comparsion with Fig. 7.
\label{hifibo1}}
\end{figure*}

\begin{figure*}[t]
\centerline{\includegraphics[width=3.5in]{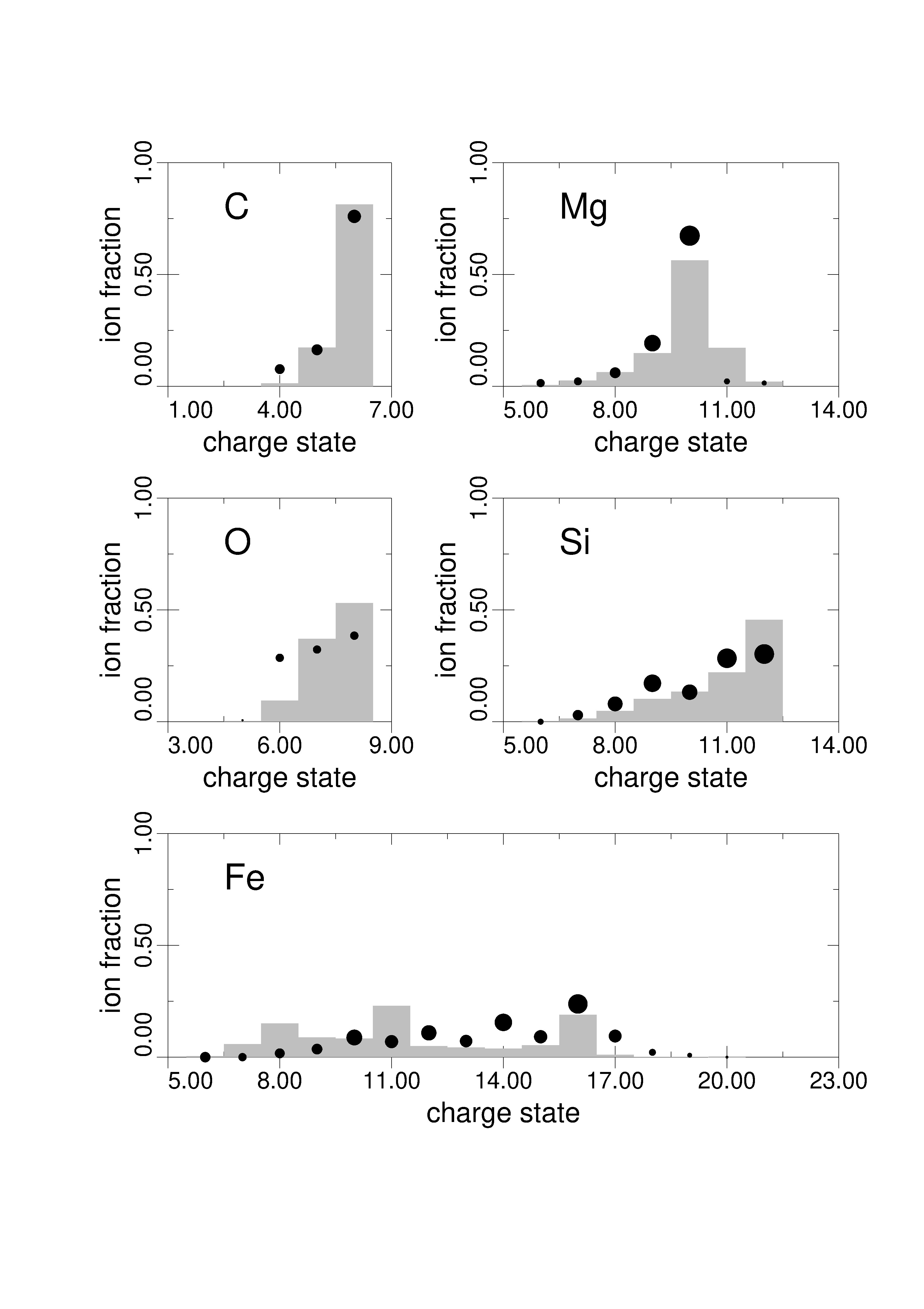}
\includegraphics[width=3.5in]{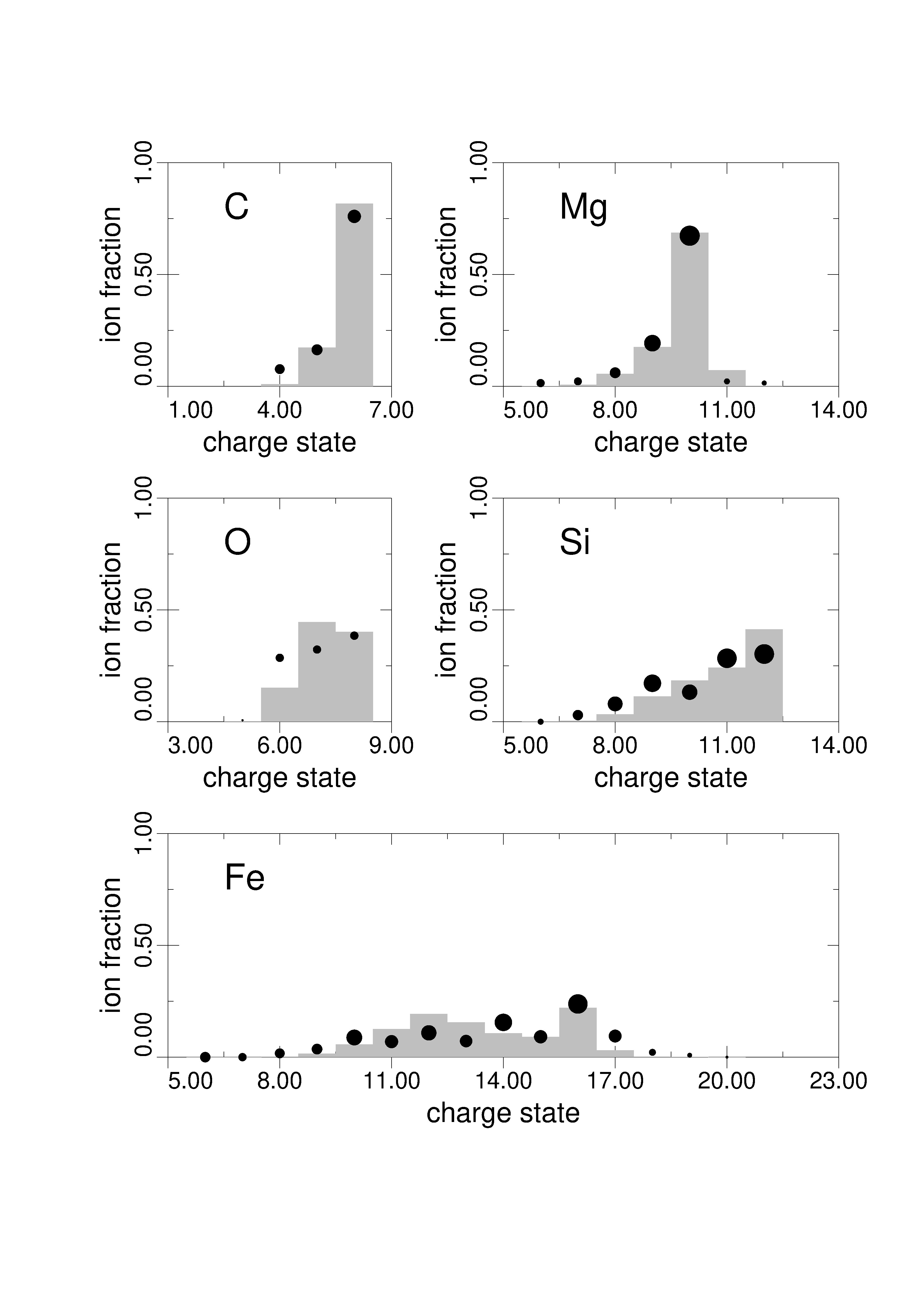}}
\vspace{-0.5true in}
\caption{Left: Final charge state distributions for C, O, Mg, Si and Fe for the 2011 February 15
CME ejecta epoch 25, assuming $\kappa = 10^6$ electron distributions. The model is shown as a histogram, and the data are shown as filled
black circles, with 2$\sigma$ uncertainties from the level 2 ACE data given by the radius. Right: The same but for $\kappa = 3.5$ electron distributions.  \label{hificmecs1}}
\end{figure*}

\subsection{2011 February 15 Event}
This follows the approach taken above, but the CME and charge states evolutions are quite different, and so this event offers an interesting contrast
with the analysis above. We start the Lagrangian plasma element at 1.07 R$_{\sun}$, with speed 75 km s$^{-1}$, density $5\times 10^9$ cm$^{-3}$ and temperature $9\times 10^6$ K. This temperatures is higher than that assumed above, and
implies some pre-eruption CME heating. This presumably happens through reconnection, in conditions (i.e. high Lundquist number) that yield much more
heat than kinetic energy. The higher pre-eruption temperature of $9\times 10^6$ K recalls the models of \citet{gruesbeck11}.  The erupting plasma in this case is essentially recombining, and gives qualitatively different charge state fractions patterns to continually heated ejecta. 

The evolution of charge state distributions and the final charge state distribution are shown for epoch 25 from Table 4, in a similar manner in Figures 6 and 7, similarly to Figures 4 and 5. The rapid acceleration and high density in this CME leads to strong cooling in the initial phase of the evolution. This faster acceleration means that it reaches it final speed, again about 1000 km s$^{-1}$, much earlier than the 2010 April 03 event. Hence all the CME heating
happens early, and the charge states continue to evolve once the heating has stopped, and the electron distribution relaxes back towards a Maxwellian. Even so, because the initial heating takes the temperature up to about $9\times 10^6$ K and being flare heating is assumed to heat the electrons into a $\kappa$ distribution, this imprint remains in some of the charge state distributions tabulated in Table \ref{febcme}, which  shows, similarly to Table \ref{aprcme}, the results of simulations compared to data for a similar range of $\kappa$ values. 
In Table \ref{febcme}, epochs 9 - 15 and 25 - 31 show a preference for $\kappa \sim 2.6-4.0$, epochs 19-22 favor $\kappa > 4.0$, while 12 and 18 show no preference. Other epochs not shown typically also show no preference for $\kappa$, and generally with worse fits than presented here.

\begin{figure*}[t]
\centerline{\includegraphics[width=3.5in]{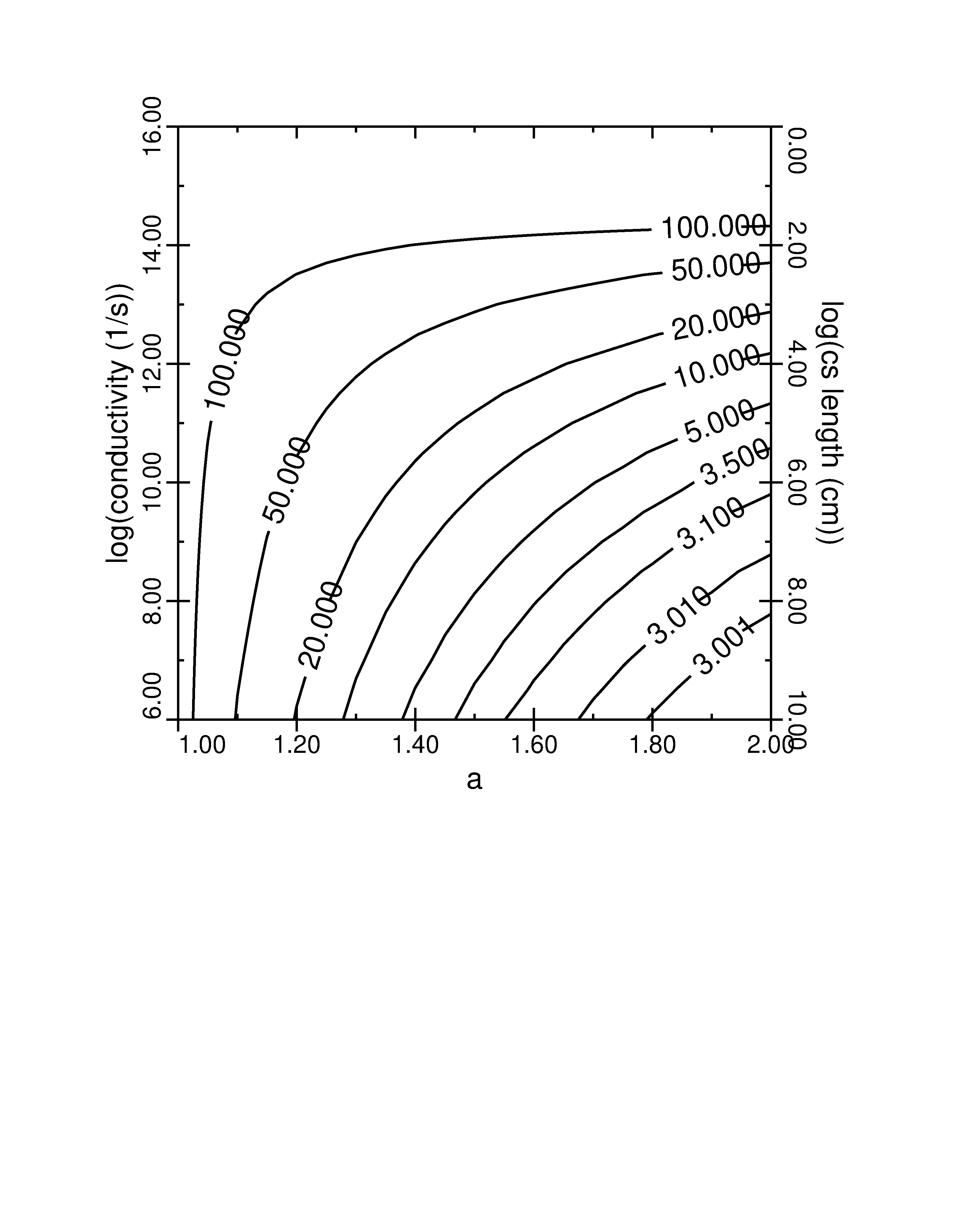}\includegraphics[width=3.5in]{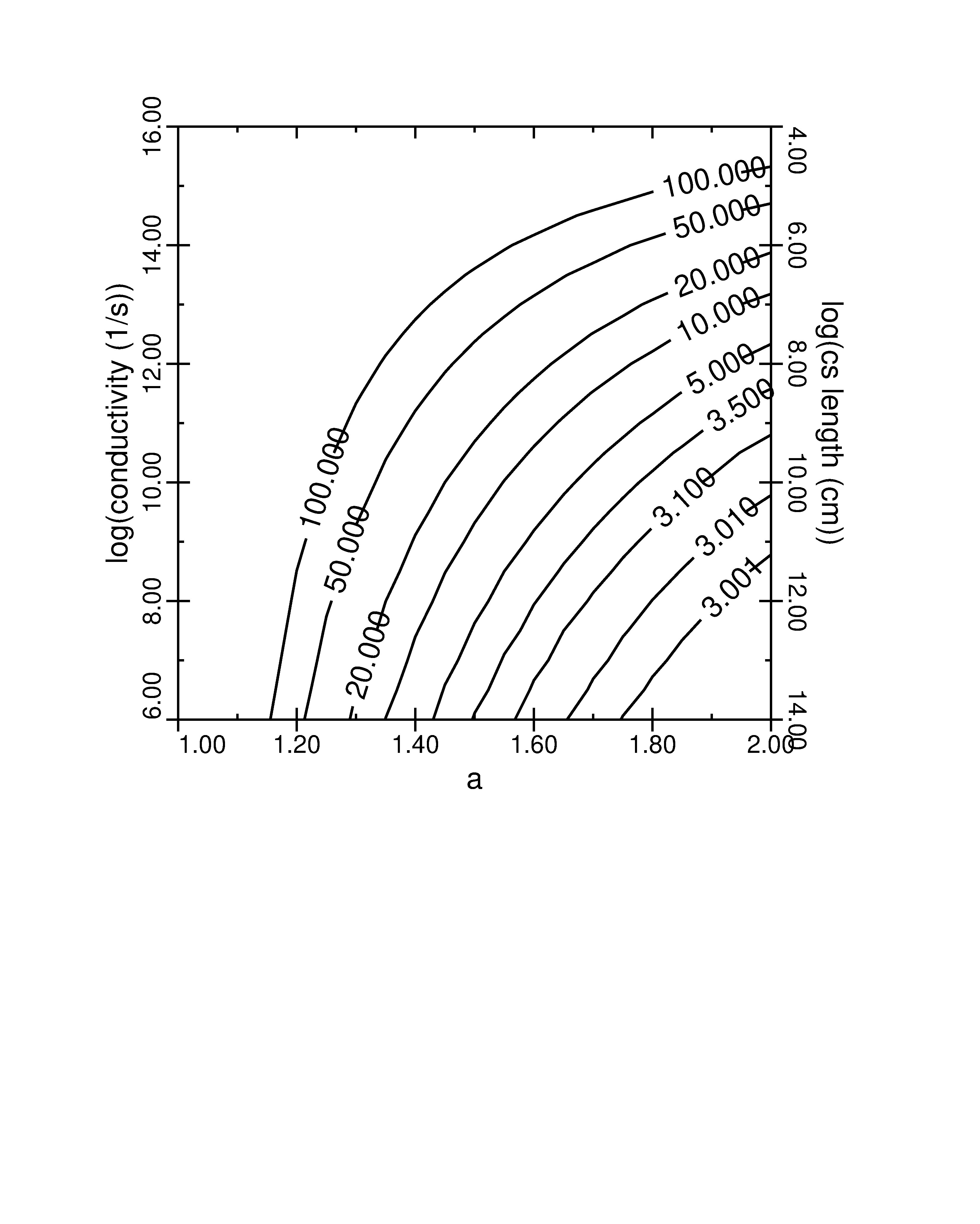}}
\vspace{-1.5true in}
\caption{Contours of electron velocity distribution function power law index $\gamma = 2\kappa$ in $\sigma - a$ space, for conductivity 
$\sigma$ and turbulent spectrum $k_{||}^{-a}$. Lundquist number $S\sim 10^4$ is assumed on the left, $S\sim 10^8$ on the right, and
$L\sigma\sim S\times 10^{12}$ cm s$^{-1}$. \label{contours}}
\end{figure*}

\section{Discussion}
We are working within a picture where reconnection current sheets with Lundquist number $S> S_c\sim 10^4$ break up into magnetic islands (plasmoids). As these plasmoids merge, by further reconnection, and contract, electrons are heated by a Fermi mechanism. This secondary reconnection, although at lower Lundquist number, may also be plasmoid unstable if $S>S_c$ still holds, and this plasmoid production, merging and electron heating will continue until $S < S_c$. Exactly what scale this occurs at is unclear. \citet{song12} observe magnetic islands appearing and merging at heights 4.73 - 7.20 R$_{\sun}$ with size
$\sim 10^{11}$ cm. Closer to the Sun, \citet{liu13} and \citet{lu22} see magnetic islands forming 0.2 - 0.25 R$_{\sun}$ above the solar surface with diameters
$\sim 2\times 10^9$ cm during the 2012 July 19 flare. Structures with these dimensions require an anomalous electrical conductivity for a Lundquist number $S\sim 10^4$. \citet{bemporad08} has previously argued for anomalous conductivities of the order implied on the basis
of observations of the widths of current sheets. Tables 1 and 2 give value of the Lundquist number based on such estimates of anomalous conductivity for the two CMEs of interest here.

Our analysis allows for some quantitative statements of such things.
The ``fitted'' values of $\kappa = \gamma /2$ compare well with those determined in flares by other methods. Analyses of hard X-ray and microwave emissions associated with downgoing accelerated electrons in the 2017 September 10 flare indicate $\kappa\simeq 2.5-4$ \citep[$\gamma = 5-8$][]{fleishman22,li22} with harder spectra coming earlier in the flare when the magnetic field, and presumably also the Lundquist number, were higher. That magnetic field destruction powers the flare is confirmed by \citet{fleishman20}, who measure the evolution of the magnetic field during the event from microwave emission, with electron jets being observed by \citet{chen18}. This observation, and the active region inferences of \citet{delzanna22} concern sunward directed electrons. For anti-sunward directed electrons, as treated in this paper, \citet{petersen21} report similar spectra in flares observed by \citet{klassen16} using the STEREO Solar Electron and Proton Telescope. These various electron spectral indices can be compared
with values to be expected from equation 9. We assume the accelerated electrons
are scattered by parallel propagating waves with energy spectrum $\propto k_{||}^{-a}$ where typically $1 < a < 2$.
In equation 9 we write $\nu = \Omega _e\pi/4 \times \left(k_{||}/k_m\right)^{-a}$ \citep[following][for scattering of ions]{laming14} where $k_m\simeq 2S^{3/8}/L$ is the wavevector where 
plasmoid growth is maximized \citep[e.g.][]{Loureiro07}, and at maximum growth we assume $\delta B \sim B$. Then
\begin{equation}
\nu = \frac{\pi}{4}\Omega _e\left(\frac{\Omega _e}{v_{||}}\frac{L}{ 2S^{3/8}}\right)^{-a} = 
\frac{\pi}{4}\Omega _e\left(\frac{\omega _{pe}}{c}\frac{V_A}{v_{||}}\sqrt{\frac{m_i}{m_e}}\frac{L}{2S^{3/8}}\right)^{-a}
\end {equation}
since $\Omega _i = \omega _{pi}V_A/c$ and $\Omega _e=\omega _{pe}V_A/c \times \sqrt{m_i/m_e}$, and equation 8 becomes
\begin{eqnarray}
\gamma &=& \frac{3}{2}+\sqrt{\frac{9}{4}+\frac{15\pi}{2}\left(\frac{\omega _{pe}}{c}\sqrt{\frac{m_i}{m_e}}\right)^{1-a}\frac{V_A^2}{v_{in}^2}
\left(\frac{v_{||}}{V_A}\right)^a\frac{\delta _{cs}^2}{L}\frac{S^{3a/8}}{\left(L/2\right)^a}}\cr
&=& \frac{3}{2}+\sqrt{\frac{9}{4} +23.56\left(\frac{\sqrt{n_e}}{ 12422}\right)^{1-a}\left(\frac{v_{||}}{V_A}\right)^a
\frac{S^{3a/8}}{\left(L/2\right)^{a-1}}}.
\end{eqnarray}
where we have assumed $\left(V_A/ v_{in}\right)^2\simeq S$ in the final step. Figure 8 plots contours of $\gamma$ in the $a$-conductivity space assuming $S\sim 10^4$ (left panel) and $S\sim 10^8$ (right panel). For a given $S$, the product of loop length and conductivity remains constant. Our inferred electron velocity distribution power law indices pick out the region towards the bottom
right hand corner of each contour plot. This matches with
\citet{zank22}, who find $a\simeq 1.5$ in the sub Alfv\'enic solar wind observed with Parker Solar Probe, and with the plasmoid dimensions referred to above. Taking $a=1.5$ and $\gamma < 8$ gives $L>10^6$ cm for $S=10^4$ and 
$L>10^{10}$cm for $S=10^8$. These estimates become $L> 10^4$cm and $L>10^7$ cm respectively for $a=2$. For
the case where $\gamma > 8$ is preferred, the reverse inequalities hold.

An independent estimate of the limiting plasmoid size comes from the requirement that the electrons are collisionless at this scale to maintain a $\kappa$ distribution. \citet{uzdensky10} argue that this is always true, but assume that the electrical conductivity remains classical. We take the electron-electron relaxation time $\tau_{ee}=1.5\times 10^{-20}v_e^3/n_e$ and demand that this be greater than the acceleration time $\tau_{acc}\simeq d/v_e$ where $d$ is the magnetic island diameter. These equations give
\begin{equation} d < 1.85\times 10^{15}\frac{\left(E/100~{\rm eV}\right)^2}{n_e}
\end{equation}
which evaluates to $3\times 10^5 - 10^6$ cm for
$n_e=2\times 10^9 - 5\times 10^9$ cm$^{-3}$. This estimate favors $S=10^4$ for parallel and perpendicular island dimensions of $l_{||}=\pi L/S^{3/8}=\pi\times 10^3S^{5/8}$ and $l_{\perp}=\delta$ in the range $ L/S^{1/2} - L/S^{1/3} =10^3S^{1/2} - 10^3S^{2/3}$, where $L=10^3S$ from results above. Similarly, the favored ratios of heating to kinetic energy increase of order 1 in Table 3 also suggests $S\sim 10^4$. In Table 4, much of the heating is already embodied in the initial temperature, so the heating to KE ratio is less clear, but is likely much higher. Note that this limit does not apply to epochs 19-22 in Table 4, where high $\kappa$ is favored by the charge state analysis.

We interpret this as meaning that higher $S$ reconnection forms magnetic islands that then merge in lower S reconnection. The process cascades until the islands are small enough to allow electrons to be accelerated without redistributing their energy through Coulomb collisions such that the non-Maxwellian distribution survives. This condition appears to be reached once $S$ reaches $10^4$, with $L\sim 10^3S$ and $\sigma \sim 10^9$ s$^{-1}$ for $a=1.5$. The last term in the square root in equation 13 is proportional to $S^{1-5a/8}$ for $L\propto S$. For $a=1.5$, this goes as $S^{1/16}$ giving harder electron spectra at lower $S$. For $a=2$, the opposite
behavior results, with this last term $\propto S^{-1/4}$ and harder electron spectra resulting for {\em increasing} $S$. A wave spectrum with $a=2$ is suggested by comparison with \citet{arnold21}. They find $\gamma = 3.3$ for $S=10^8$ and $L=10^9$ cm, which matches with $a=2$ in Figure 8 (right panel). They also find harder electron spectra at higher Lundquist number. This difference in behavior with $a$ can also be seen in Figure 8.


\section{MHD simulations of Magnetic Reconnection}

To show the formation of multiple magnetic islands in a large-scale reconnection current sheet with a high S number and subsequent reconnection of merging islands at smaller S numbers, we model a post-CME current sheet in the solar corona in a compressible MHD simulation. We start from a 2.5D Harris current sheet with initial conditions that characterize the solar corona's plasma and magnetic field and are identical to our previous work \citet{Provornikova16}. This work examined the Sweet-Parker-like reconnection current sheet formed for the global Lundquist numbers $S \lesssim 10^4 $, which may occur in the corona due to enhanced magnetic diffusivity $10^{11}-10^{12}$ cm$^2$s$^{-1}$ caused by turbulence and kinetic effects (see discussion in Sec. \ref{secAR}). Here we model reconnection with magnetic islands, which occurs at S numbers larger than the critical S number, typically $S_{crit} \sim 10^4$. We use the GAMERA (Grid Agnostic MHD for Extended Research Application) MHD code \citep{Zhang2019} to solve compressible MHD equations in the Harris current sheet setup. A predecessor of the GAMERA, the LFM (Lyon-Fedder-Mobarry) MHD code, was used to simulate in a box the magnetic reconnection process in the Earth magnetotail \citep{MerkinSitnov2016}. 

\begin{figure*}[t]
\vspace{-0.5true in}
\centerline{\includegraphics[width=7in]{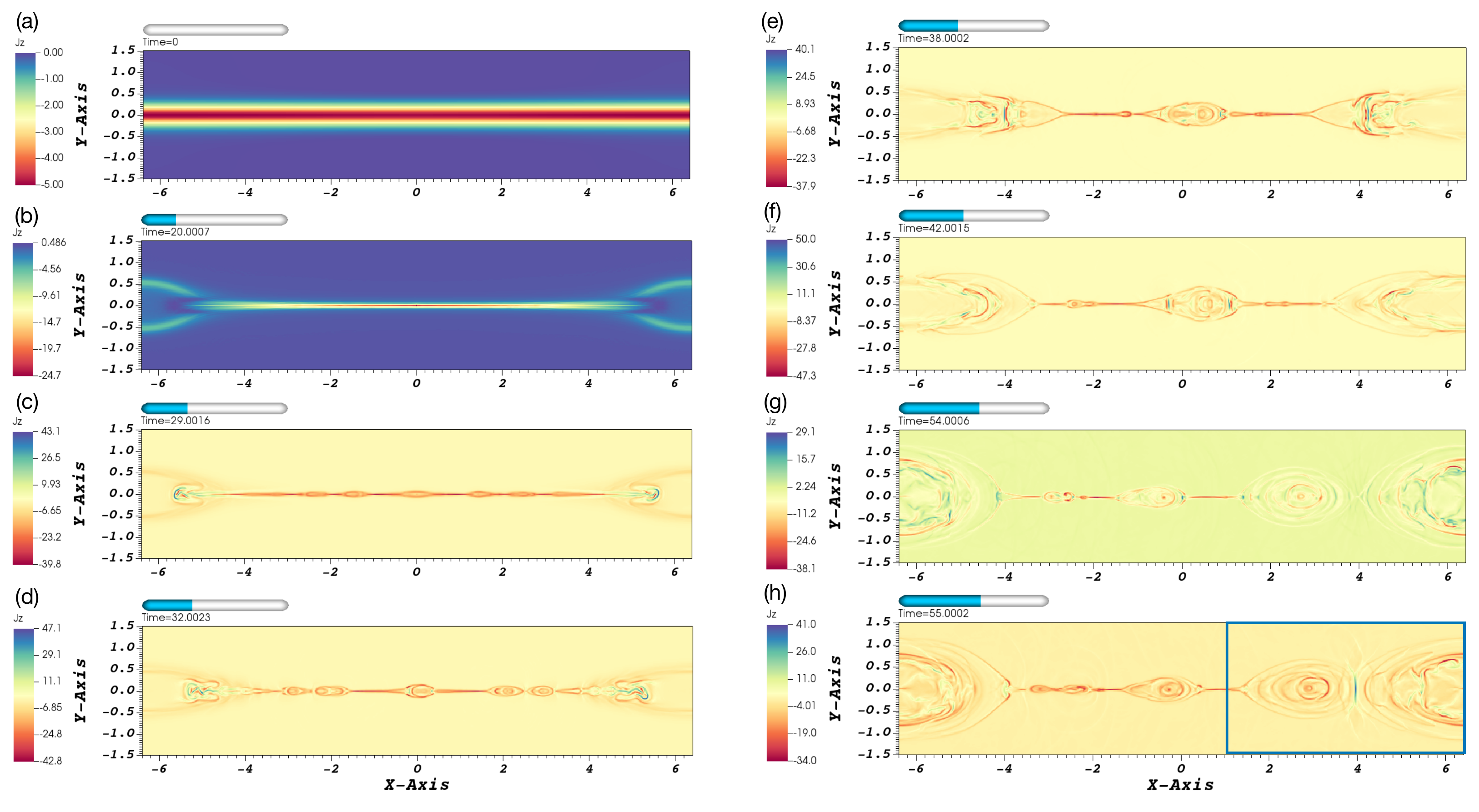}}
\vspace{0.1true in}
\centerline{\includegraphics[width=5in]{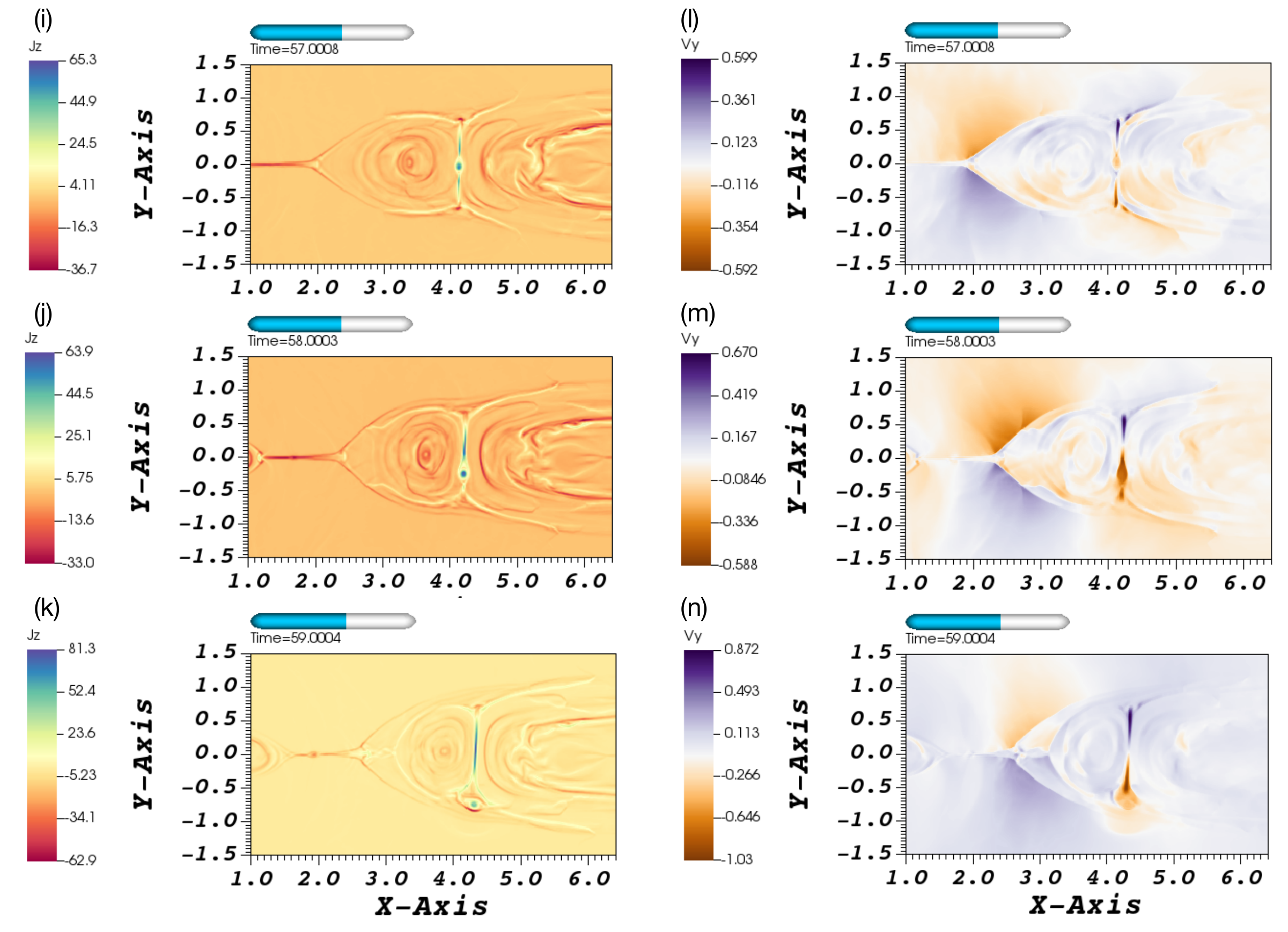}}
\caption{ GAMERA MHD simulation of magnetic reconnection in Harris current sheet representing post-CME current sheet in the solar corona. Panels (a-h) show out-of-plane current $J_z$ in the simulation domain. Panels (i-k) show out-of-plane current $J_z$ when two big islands form a secondary current sheet where reconnection with formation of smaller islands occur (blue box from panel (h)), panels (l-n) show $V_y$ velocity component indicating reconnection outflow between merging and reconnecting islands. $J_z$ and $V_y$ are shown in normalized units. Time is normalized to the Alfv\'en time. \label{plasmoids}}
\end{figure*}

The GAMERA code solves normalized MHD equations. We assume the following (normalizing) plasma and magnetic field parameters in the coronal plasma ambient to the CME current sheet within a few solar radii from the Sun. The chosen parameters are informed by MHD simulations of the global solar corona \citep{Linker1999} available for the Carrington rotation 2095 (when the eruption of April 3, 2010 CME occurred) by the Predictive Science group \footnote{\url{https://www.predsci.com/hmi/summary_plots.php}}. The plasma density is assumed to be $n=10^{10}$ cm$^{-3}$, magnetic field $B=10$ G, temperature $T=10^6$ K, and characteristic length of the system $L = 0.1R_{\sun} = 70$ Mm. The Alfv\'en speed in the reconnection inflow is $\sim 200$ km s$^{-1}$ and plasma beta parameter $\beta = p_{th}/p_{mag} = 0.07$. 

The initial conditions in the Harris current sheet are set by the magnetic field profile $B_x = B_0 tanh(y/\lambda)$ and the density $n = n_0 + (1/\beta) \cosh^{-2}(x/\lambda)$ in a simulation box $[-L_x, L_x]$, $[-L_y, L_y]$, and $[-L_z, L_z]$ with $L_x = 2\pi$, $L_y = \pi$, and $L_z = 0.1$ (length scales are normalized to $L$). The resolution of the spatial grid is $N_x =864$, $N_y = 432$, and $N_z =8$. The thickness of the current sheet $\lambda = 0.2$. Magnetic reconnection is initiated by introducing a small-amplitude disturbance of the magnetic potential $\mathbf{B} = \mathbf{z} \times \nabla \psi$ in a form $\psi = 0.1 B_0 \cos (2\pi x/L_x) \cos(\pi y /L_y)$.
Boundary conditions are set to be periodic at the left/right boundaries perpendicular to the current sheet and perfectly conducting in the top/bottom boundaries parallel to the current sheet. GAMERA reconnection simulations were performed at CISL Cheyenne system \citep{cheyenne}.

The disturbance initiates a magnetic reconnection process in the Harris current sheet (Figure \ref{plasmoids}). The reconnection current sheet becomes thinner, elongates (panel (b)), and breaks up into magnetic islands (panel (c)). The islands grow and merge, forming larger islands (panels (d-g)). When two islands are pushed to each other, the oppositely directed magnetic field with a dominant $B_y$ component in adjacent islands forms a vertical current sheet (panel (h)). We find from simulations that the ratio of a local Lundquist number in such a ``secondary'' current sheet to a global Lundquist number for the initial current sheet is $S_{local}/S_{global} \sim 0.01 - 0.1$ depending on the length of the secondary current sheet. This agrees well with the ratio expected from the analytic theory, $S_{local}=\pi LV_A/\eta S_{global}^{3/8}$, which is derived by replacing $L$ with
$2\pi/k_{max}$ where $k_{max}\simeq 2S^{3/8}/L$, the wavenumber of maximum growth.

If $S_{local}$ exceeds the critical S value, the reconnection process in a local current sheet produces small-scale islands. Panels (i-n) in Figure \ref{plasmoids} show a formation and ejection of a magnetic island in a vertical current sheet. Since reconnection occurs at a lower S, fewer magnetic islands are expected to form \citep{Bhattacharjee10} in secondary current sheets, and these islands will be smaller, allowing non-Maxwellian electron distributions a better chance of surviving. Such processes continue until the S number for current sheets forming in this cascading process of island merging reaches values below $S_{crit}$. In our cases above, only reconnection just above $S_{crit}$ is sufficiently collisionless to allow the electron $\kappa$ distribution to survive and ionize the ejecta, but this need not always be the case. We speculate that more generally, ionization of the ejecta should be characteristic of the hardest electron spectrum produced in reconnection, and depending on the spectrum of turbulence, this can be either at high or low $S$, (but still above $S_{crit}$).

Our MHD simulation of magnetic reconnection in a large-scale coronal current sheet with a high S number shows that magnetic islands coalesce and form smaller-scale secondary current sheets, which may further be disrupted depending on the S number into the smaller-scale magnetic islands. Such system evolution supports findings from the analysis of charge states that electrons are accelerated in small-scale islands formed in lower S number current sheets, as discussed in Sec. 6.

\section{Conclusions}
In this paper, we have attempted to model ion charge states in CME ejecta from the initial reconnection event to their eventual detection by instrumentation at 1 AU. 
This ``end-to-end'' analysis involves characterization of the reconnection and the MHD expansion of the CME ejecta from observations with STEREO A and B at near quadrature with the Earth and detection of the ejecta particles themselves by ACE/SWICS near the L1 Lagrange point. We find electron velocity distribution function,
characterized by a kappa distribution with values $\kappa = 2.6 - 4.0$. This is more clearly seen for certain times during the 2011 February 15 CME where the charge state evolution is dominated by recombination. Our $\kappa$ value are close to the theoretical minimum of $\kappa = 1.5$, but this value is actually significantly disfavored.
For reference, 
\citet{delzanna22} and \citet{fleishman22} also observe $\kappa = 3-5$ by different means, and \citet{li22} model $\kappa \simeq 3$ for the 2017 September 10 event, observed by \citet{fleishman22}. Applying these results to the theory of electron acceleration by merging magnetic islands in the reconnection current sheet constrains the Lundquist number to values close to its critical value of $S=10^4$. We speculate that no matter what the initial ``global'' value of $S$, so long as $S>10^4$, the merging of the islands so produced generates new current sheets at lower $S$ until they become subcritical, as in \citet{uzdensky10} and \citet{huang13}. If any of these islands form and merge under conditions that render the electrons collisionless, non-Maxwellian electron velocity distribution functions, modeled here as $\kappa$ distributions, may result in subsequent effects in the ionization of the CME ejecta
and distortion of the charge state distribution from that expected with Maxwellian electrons, subsequently detected after freeze in at 1 AU.

\acknowledgments
This work has been supported by NASA HSR Grants NNH22OB102 and 80HQTR19T0064, and by Basic Research Funds of the Office of Naval Research.
E.P. acknowledges support from the NASA Heliophysics Supporting Research program (contract 80NSSC19K0856). We acknowledge high-performance computing support from the NSF-sponsored Cheyenne system provided by NCAR's Computational and Information Systems Laboratory (CISL). The authors thank Kareem Sorathia, Slava Merkin, and Harry Arnold for helpful discussions.

\end{document}